\documentclass[preprint2]{aastex}
\def\bcga{{A1139-BCG}}
\def\bcgb{{A2589-BCG}}
\def\umu{{mag arcsec$^{-2}$}}

\shorttitle{pCMDs of BCGs in A1139 and A2589}

\shortauthors{Lee et al.}
\def\simlt{\lower.5ex\hbox{$\; \buildrel < \over \sim \;$}}
\def\simgt{\lower.5ex\hbox{$\; \buildrel > \over \sim \;$}}

\begin{document}

\title{ Pixel Color-Magnitude Diagram Analysis of the Brightest Cluster Galaxies in Dynamically Young and Old Clusters, Abell~1139 and Abell~2589}

\author{Joon Hyeop Lee$^{1,2}$, Sree Oh$^{3}$, Hyunjin Jeong$^{1}$, Sukyoung K. Yi$^{3}$, Jaemann Kyeong$^{1}$, and Byeong-Gon Park$^{1,2}$}
\affil{$^1$ Korea Astronomy and Space Science Institute, Daejeon 305-348, Republic of Korea\\
$^2$ University of Science and Technology, Daejeon 305-350, Republic of Korea\\
$^3$ Department of Astronomy and Yonsei University Observatory, Yonsei University, Seoul 120-749, Republic of Korea}

\email{jhl@kasi.re.kr}

\begin{abstract}
As a case study to understand the coevolution of Brightest Cluster Galaxies (BCGs) and their host clusters, we investigate the BCGs in dynamically young and old clusters, Abell~1139 (A1139) and Abell~2589 (A2589). We analyze the pixel color-magnitude diagrams (pCMDs) using deep $g$- and $r$-band images, obtained from the Canada-France-Hawaii Telescope observations. After masking foreground/background objects and smoothing pixels in consideration of the observational seeing size, detailed pCMD features are compared between the two BCGs. (1) While the overall shapes of the pCMDs are similar to those of typical early-type galaxies, the A2589-BCG tends to have redder mean pixel color and smaller pixel color deviation at given surface brightness than the A1139-BCG. (2) The mean pixel color distribution as a function of pixel surface brightness (pCMD \emph{backbone}) indicates that the A2589-BCG formed a larger central body ($\sim2.0$~kpc in radius) by major dry mergers at an early epoch than the A1139-BCG (a central body $\sim1.3$~kpc in radius), while they have grown commonly by subsequent minor mergers. (3) The spatial distributions of the pCMD outliers reveal that the A1139-BCG experienced considerable tidal events more recently than the A2589-BCG, whereas the A2589-BCG has an asymmetric compact core possibly resulting from major dry merger at an early epoch.  (4) The A2589-BCG shows a very large faint-to-bright pixel number ratio compared to early-type non-BCGs, whereas the ratio for the A1139-BCG is not distinctively large. These results are consistent with the idea that the BCG in the dynamically older cluster (A2589) formed earlier and is relaxed better.
\end{abstract}


\section{INTRODUCTION}

Brightest cluster galaxies (BCGs) are very special objects, in which much information is stored about how baryons are assembled into massive galaxies. The overall appearances of most BCGs are similar to giant elliptical galaxies, but they are known to be different in several detailed properties, such as the mass-size relation, mass-to-light ratio, and spatial distribution of stellar populations  \citep{von07,lou12}. 
Owing to the high luminosities as well as the scientific importance of BCGs, they have been excellent targets in observations and subsequent theoretical studies for a long time. Already in 1970s -- 1980s, three classical scenarios of BCG formation were established: galactic cannibalism \citep{ost75,kor84}, primordial origin \citep{mer83,mer84}, and accretion of cooling flow \citep{cow77,fab77}.
Today, however, many studies show that the observed properties of BCGs are so complicated and various that any single scenario can not fully explain every aspect of BCGs \citep{bro07b,lou12,jim13}.
That is, BCGs may not be a homogeneous galaxy population and thus they may have diverse histories of star formation and mass assembly.

One of the strong suspects that cause such diversity is environment.
It is well known that the properties of galaxies are considerably influenced by their environments \citep[e.g.,][]{dre80,par07,lee10}.
Particularly, cluster environments tend to affect the properties of their member galaxies in various ways, such as ram-pressure stripping, harassment, strangulation and hydrodynamic interactions \citep[e.g.,][]{gun72,lar80,moo96,mcc08,par09}. Such effects can significantly change the properties of galaxies or small groups that fall into a cluster \citep[e.g.,][]{oem09,lee14,gu16,lee16}. However, the environmental effects on BCGs by their host clusters may not necessarily be the same as those on non-BCGs, because the formation process of a BCG is believed to be different from non-BCGs although its details are not perfectly understood yet \citep{bat07,von07,fas10,wen15}.

Because BCGs are the most dominating galaxies usually at the centers of clusters, it is natural to expect that their formation is connected with the formation of their host clusters. Several recent observational studies support such expectation.
For example, \citet{bel16} presented a positive correlation in mass between BCGs and their host clusters, although the scatter is large. This may be a simple but important evidence of the BCG-cluster coevolution.
\citet{lou12} reported that the metallicity gradient of a BCG depends on the distance from the BCG to the X-ray peak of its host cluster. This implies that the mass assembly of a BCG is influenced by the dynamical stage of the host cluster.
\citet{lau14} found that the photometric curve of growth of a BCG depends on the spatial offset from the host cluster center. From this finding, they inferred that the inner portions of the BCGs may be formed outside the cluster, but interactions in the center of the galaxy cluster extend the envelopes of the BCGs.
Meanwhile, \citet{has14} showed that morphologically disturbed clusters tend to harbor brighter BCGs, arguing that the early collapse may not be the only major mechanism to drive the BCG formation and evolution.

Despite the possible connection between BCGs and their host clusters revealed by several previous studies, the properties of BCGs as a function of their host cluster environments are not sufficiently comprehended yet. The best approach to understand their coevolution is to compare BCGs hosted by galaxy clusters in various evolutionary stages; for example, relaxed versus unrelaxed, fossil \citep{jon03} versus non-fossil, and cool-core \citep{jon84} versus non-cool-core clusters.
However, this kind of comparisons require large observational resources, not only because a large sample of clusters is necessary (with sufficient depth and spatial coverage to reliably estimate the properties of host clusters), but also because BCGs are very large targets compared to normal galaxies.
The high luminosities of BCGs make observations easier, but the large angular sizes of BCGs at low redshifts make it difficult to cover their entire bodies with spectroscopy. That is, nearby BCGs are excellent targets for imaging, but the spectroscopy for their entire bodies is not easy work.
Thus, while we try to secure spatially-resolved spectroscopic information of nearby BCGs, the efforts to fully utilize their image data are also necessary.

One of the latter efforts is the pixel analysis. Since early 21st century, some researchers tried to make use of the photometric information of individual pixels from image data sets. \citet{con03} estimated star formation histories of individual pixels in galaxies using the \emph{pixel-z} technique, and \citet{joh05} conducted pixel-by-pixel SED fitting  with $BVRIK$-bands images for a galaxy hosting a compact radio source.
More recently, the ages, star formation rates, dust contents and metallicities of 45,000 galaxies were measured in this method \citep{wel08,wel09}, and \citet{wij10} investigated the spatial distribution of pixel quantities of the Sloan Digital Sky Survey \citep[SDSS;][]{yor00} galaxies  as a function of galaxy morphology.
The pixel-z technique is a powerful method among photometric approaches, but it also has shortcomings. Even if we set aside the fact that the pixel-z technique depends on spectral energy distribution (SED) templates and fitting algorithms, it requires images in many (at least four) bands for high reliability as all photometric SED fitting methods do. Although this may require less observational resources than spectroscopy covering the entire body of a BCG, it is still not easy to secure wide-field images in diverse bands, the central wavelengths of which should be sufficiently separated for reliable SED fitting.

A simpler way is the pixel color-magnitude or color-color diagram analysis.
\citet{kas03} built the pixel-by-pixel maps of stellar populations and dust extinction for merging galaxies NGC 4038/4039, using the images in the $BVK$ bands. They approximated age and reddening of each pixel in the galaxies by comparing their pixel color-color diagrams (pCCDs) with population synthesis models.
In a similar way, \citet{deg03} estimated the star formation histories of two interacting galaxies, using their pixel color-magnitude diagrams (pCMDs) and pCCDs. They compared the spatial distributions of pixels in different domains of the pCMD and pCCD, finding significant influence of galaxy interactions on stellar populations. 
From the estimation of resolved mass maps using the optical versus near-infrared pCCD, \citet{zib09} showed that the pixel analysis is stronger than integrated photometry in estimating the stellar mass distributions of galaxies.
\citet{lan07} investigated the pCMDs of 69 nearby galaxies with various morphological types, characterizing several pCMD features for given galaxy morphology. For example, early-type galaxies show \emph{prime sequences} but some of them have \emph{red hook} features possibly originating from dust. The pCMDs of lenticular galaxies appear to be similar to those of elliptical galaxies but their prime sequences tend to be more deviated. In the pCMDs of several face-on spirals, \emph{inverse-L} features are found.
More recently, \citet{lee11,lee12} applied the pCMD and pCCD analysis method to M51, showing that those methods are very useful to understand many details about stellar populations and their spatial distributions for individual galaxies.

In this paper, we present our case study on two nearby BCGs hosted by galaxy clusters in different evolutionary stages, Abell~1139 (A1139) and Abell~2589 (A2589), using the pCMD analysis method. The goals of this paper are (1) to quantify the detailed properties of the two target BCGs from the pixel analysis, (2) to investigate the difference between the two BCGs, possibly related with their host clusters, and (3) to compare the BCGs with non-BCGs in the same clusters. Through these investigations, we intend to enhance our understanding about the coevolution of BCGs and their host clusters.
The outline of this paper is as follows. Section~2 describes the targets and observations. Section~3 shows our pixel analysis procedure. The results are presented in Section~4, and their implication is discussed in Section~5. In Section~6, the paper is concluded. Throughout this paper, we adopt the cosmological parameters: $h=0.7$, $\Omega_{\Lambda}=0.7$, and $\Omega_{M}=0.3$.

\section{TARGETS AND OBSERVATIONS}\label{data}

\begin{figure*}[t]
\centering
\includegraphics[width=0.45\textwidth]{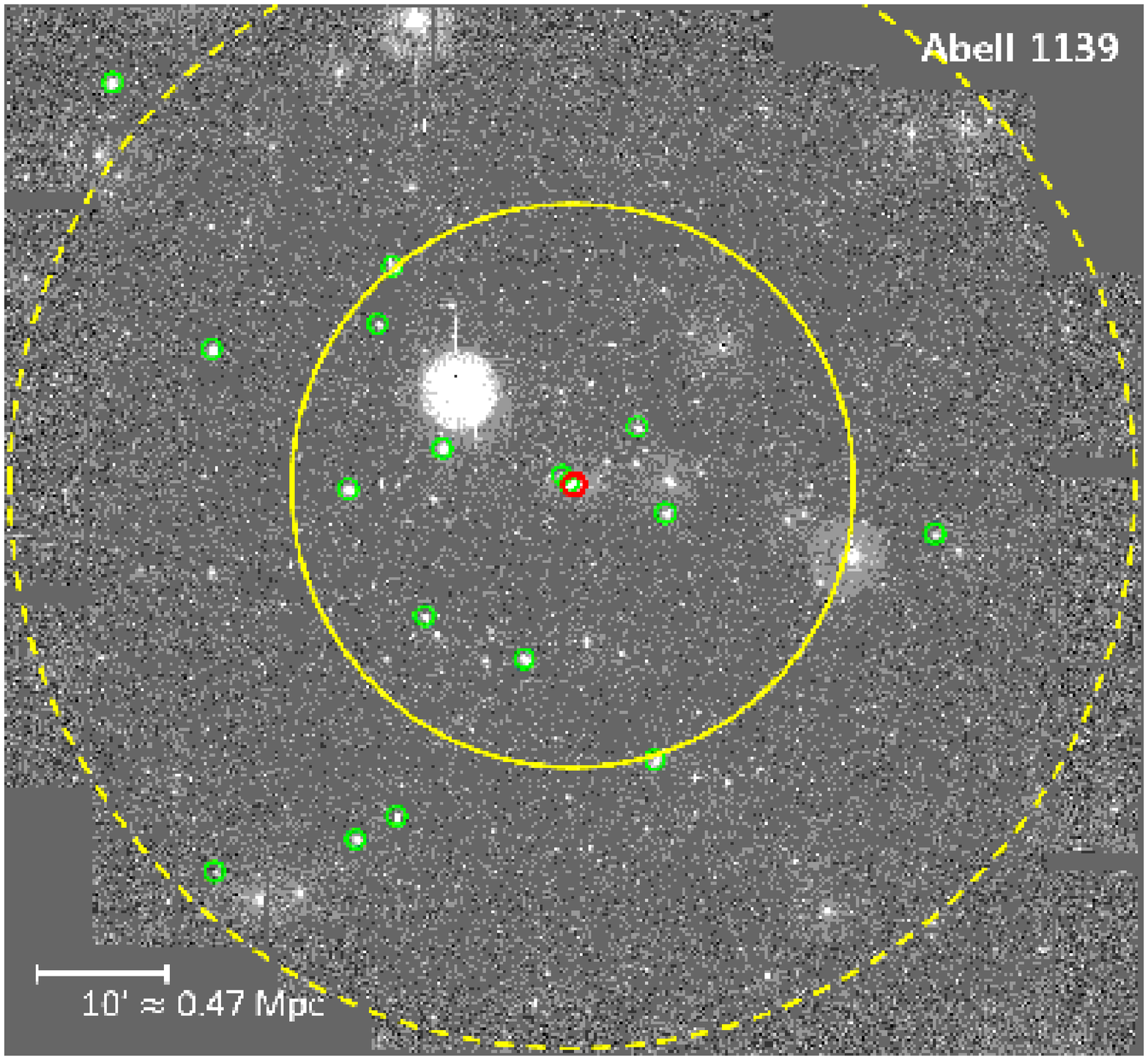}
\includegraphics[width=0.45\textwidth]{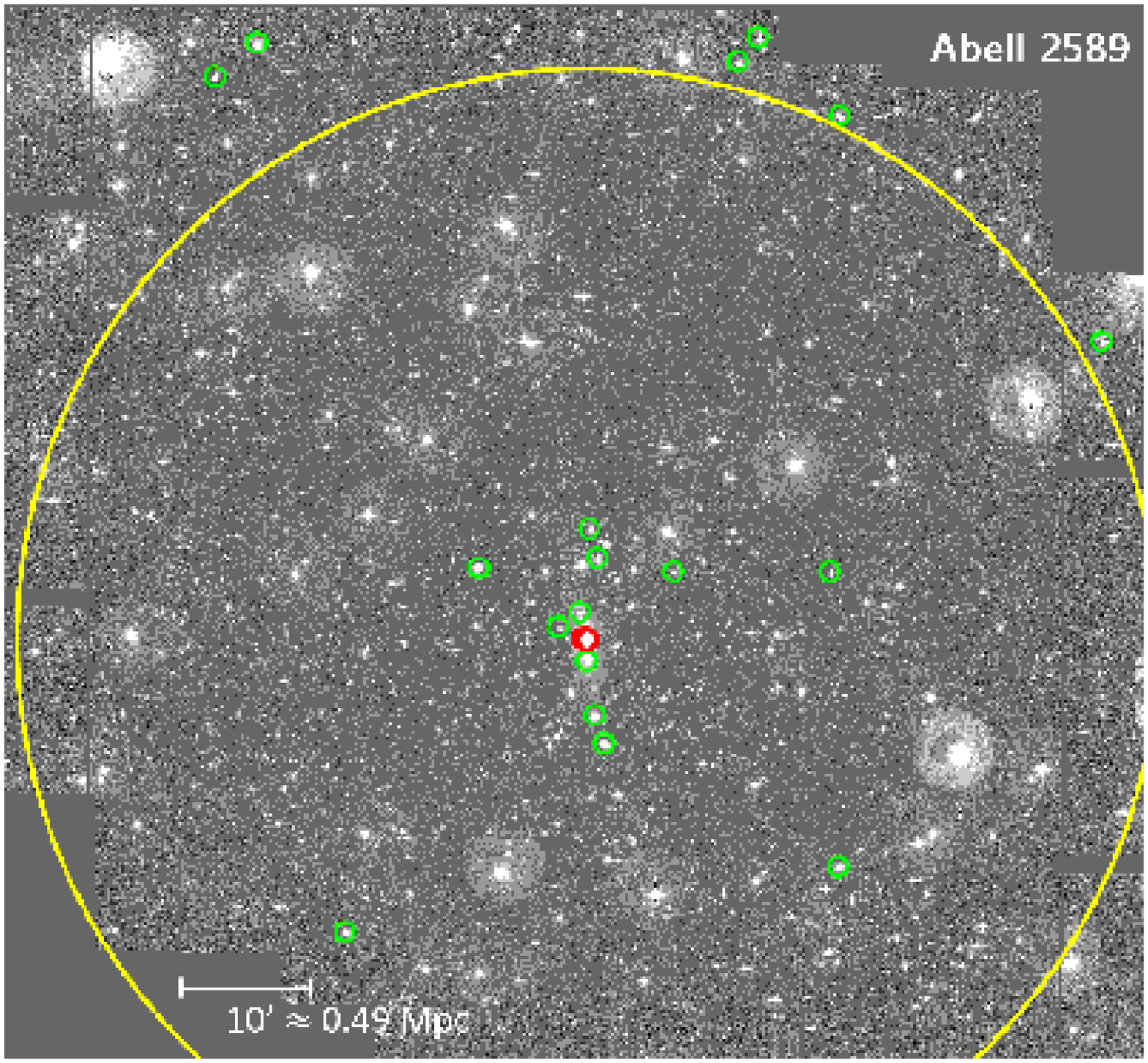}
\caption{The stacked $r$-band images of the two target clusters A1139 and A2589. In each cluster, the red circle indicates the brightest cluster galaxy (BCG) and the green circles are the bright member galaxies ($M_r\le -21$ and $|\Delta v_{rev}| \le 3 \sigma_{cl}$). The solid yellow circles show the virial radii ($R_{200}$) and the dashed yellow circle displays the $2 \times R_{200}$ of A1139. \label{targets}}
\end{figure*}

We obtained our data set from the queue observations using the 3.6-meter Canada-France-Hawaii Telescope (CFHT) in 2012 -- 2013, as a part of the KASI-Yonsei Deep Imaging Survey of Clusters (KYDISC) project, which targets 18 nearby ($0.01\le z \le 0.20$) rich clusters to understand the formation histories of their member galaxies.
Among the KYDISC clusters, we investigate two galaxy clusters at very similarly low redshifts ($z\sim0.04$) but in quite different evolutionary stages: A1139 (young) and A2589 (old). 
\citet{hwa07} argued that A1139 is probably rotating or merging; in other words, A1139 is a dynamically young cluster. On the other hand, A2589 is known to be a well-relaxed cluster \citep{buo96,bau00,buo04,liu11}.
The basic information of the two target clusters and their BCGs are summarized in Table~\ref{clinfo}.

The selection of targets at similar redshifts provides two technical advantages. First, the conversion factors from angular size to physical scale are similar: 0.79 kpc per arcsec for A1139 and 0.82 kpc per arcsec A2589. This means that the target images have similar spatial resolutions, which is important for fair comparison of their pCMDs because the image resolution significantly influences the detailed features in pCMDs \citep{lee11,lee12}. It is possible to rescale the images when the resolutions of targets are significantly different, but it will cause data loss for the target at lower redshift.
Second, the central wavelengths at the rest frame are also similar. Since the observed images of targets at different redshifts have different central wavelengths at the rest frame, the interpretation of their pCMDs becomes complicated, unless k-corrections are conducted. However, if the number of image bands is not sufficiently large, the k-correction is less reliable. Moreover, because we need to conduct k-corrections for individual pixels often with low signal-to-noise ratio (S/N), the reliability of the k-correction will be even lower.
By selecting target clusters at similarly low redshifts, we do not need to risk these difficulties.
For higher statistical significance, it is worth trying to overcome these technical problems and comparing a large sample of clusters at various redshifts, but the intensive comparisons of two BCGs at similarly low redshifts will be a good starting point of the comprehensive studies on the BCG-cluster coevolution.

\begin{deluxetable}{lcc}
\tablenum{1} \tablecolumns{3} \tablecaption{ Basic Information of the Target Clusters and Their BCGs} \tablewidth{0pt}
\tablehead{ & Abell 1139 & Abell 2589 }
\startdata
Right Ascension (RA) & 10$^{h}$ 58$^{m}$ 04.3$^{s}$ $^{(a)}$ & 23$^{h}$ 23$^{m}$ 53.5$^{s}$ $^{(b)}$ \\
Declination (Dec) & +01$^{d}$ 29$^{m}$ 56$^{s}$ $^{(a)}$ & +16$^{d}$ 48$^{m}$ 32$^{s}$ $^{(b)}$ \\
Redshift$^{(c)}$ & 0.0398 &  0.0414 \\
Bautz-Morgan Type$^{(d)}$ & III & I  \\
L$_{0.1-2.4 keV}$ (ROSAT$^{(e)}$) & $(2.0\pm0.3)\times10^{43}$ erg s$^{-1}$ $^{(f)}$ &  $(9.9\pm1.1)\times10^{43}$ erg s$^{-1}$ $^{(g)}$ \\
Velocity Dispersion$^{(h)}$ & 433 km s$^{-1}$ & 879 km s$^{-1}$ \\
Virial Radius ($R_{200}$)$^{(h)}$ & 1.0 Mpc & 2.1 Mpc \\
\hline
BCG Name &  UGC 06057 NED02 & NGC 7647 \\
BCG RA & 10$^{h}$ 58$^{m}$ 11.0$^{s}$ $^{(i)}$ & 23$^{h}$ 23$^{m}$ 57.4$^{s}$ $^{(j)}$ \\
BCG Dec & +01$^{d}$ 36$^{m}$ 16$^{s}$ $^{(i)}$ & +16$^{d}$ 46$^{m}$ 38$^{s}$ $^{(j)}$ \\
BCG Redshift & 0.0385$^{(k)}$ & 0.0411$^{(c)}$ \\
$m - M$ & 36.23 & 36.32 \\
BCG FUV (GALEX$^{(m)}$) & $19.97\pm0.09$ & $20.32\pm0.24$ \\
BCG NUV(GALEX) & $19.43\pm0.05$ & $19.44\pm0.10$ \\
BCG $g$-band (SDSS) & $14.46\pm0.00$ & $13.56\pm0.00$\\
BCG $r$-band (SDSS) & $13.55\pm0.00$ & $12.65\pm0.00$\\
BCG $K_s$-band (2MASS$^{(n)}$) & $10.79\pm0.05$ & $10.00\pm0.06$ \\
\enddata
\tablecomments{(a) \citet{abe89}. (b) \citet{pif11}. (c) \citet{str99}. (d) \citet{bau70}. (e) R{\"o}ntgensatellit \citep{vog99}.  (f) \citet{pop04}. (g) \citet{sha09}. (h) Oh et al.~(in preparation). (i) \citet{wen09}. (j) \citet{cav09}. (k) \citet{smi00}.  (l) Galaxy Evolution Explorer \citep{mar05}. (m) Two Micron All Sky Survey \citep{skr06}. Note that the total magnitudes may significantly vary according to the definition of the spatial extent and that it is not easy to exactly determine the spatial extent of a BCG because of its outer envelope and intracluster light.}
\label{clinfo}
\end{deluxetable}

\begin{figure*}[!t]
\centering
\includegraphics[width=0.43\textwidth]{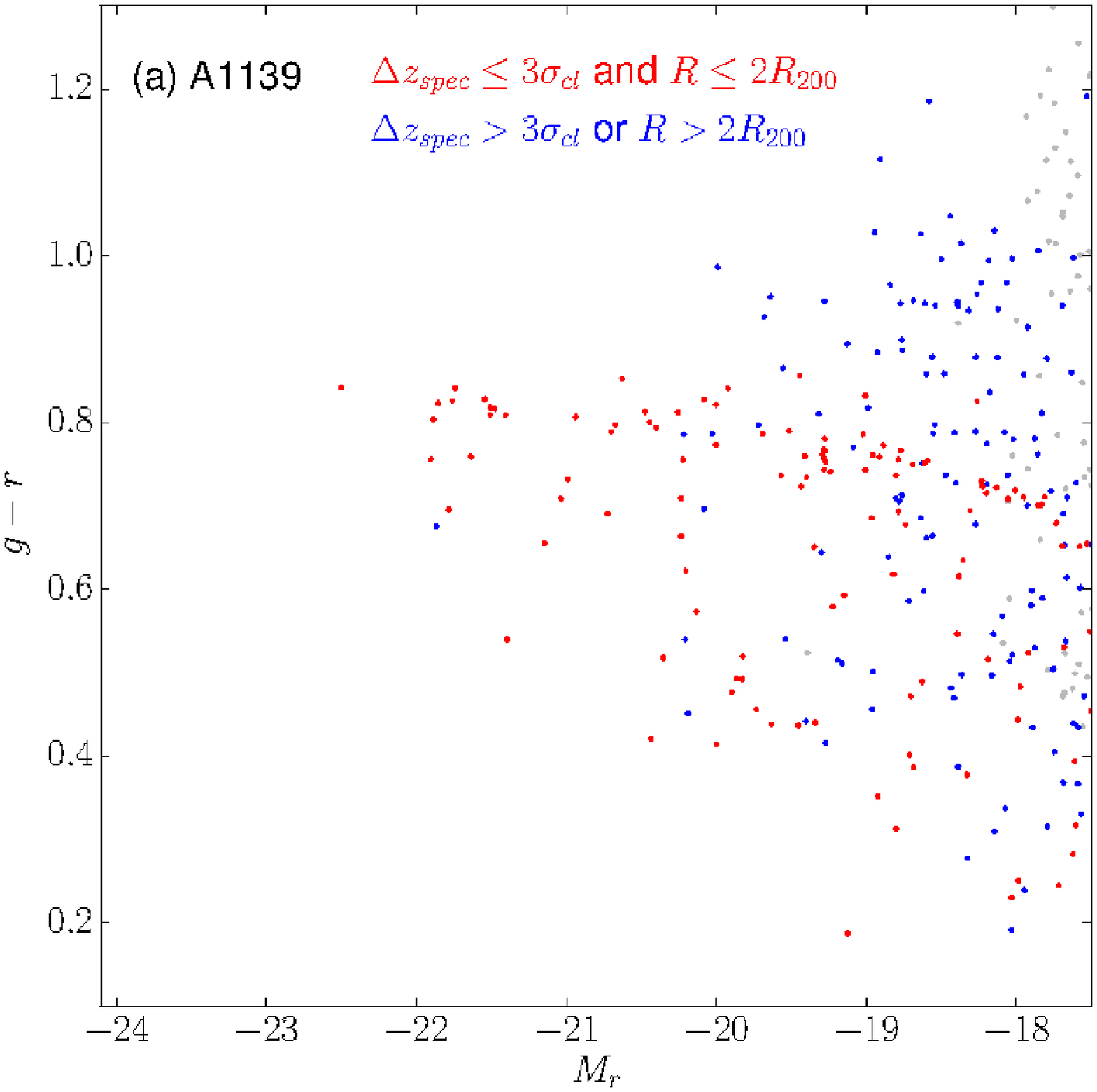}
\includegraphics[width=0.43\textwidth]{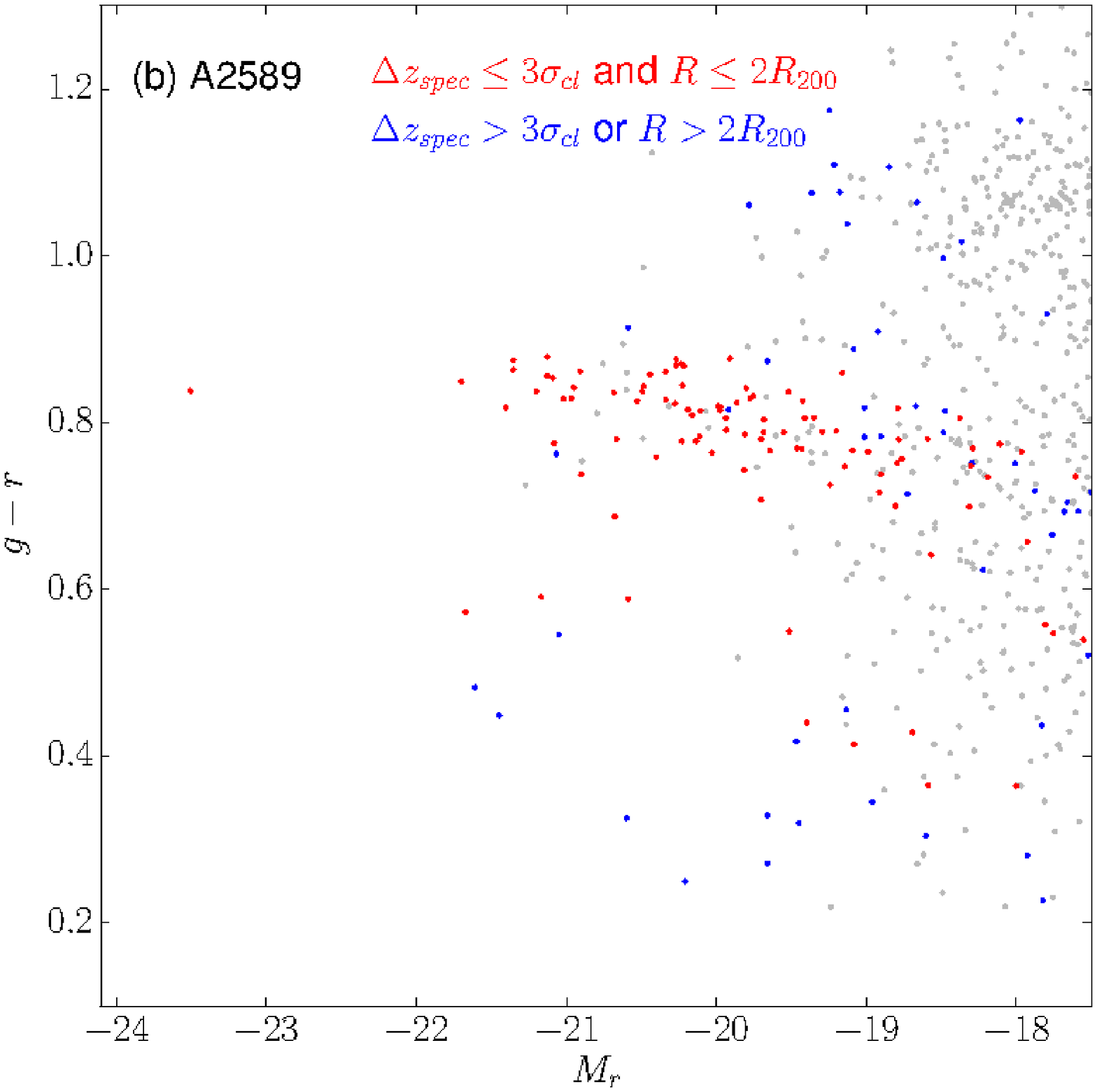}
\caption{Galaxy color-magnitude diagrams of the target clusters. The spectroscopically-selected cluster members ($|\Delta v_{rec}| \le 3 \sigma_{cl}$ and $R \le 2 R_{200}$; red dots) and non-members ($|\Delta v_{rec}| > 3 \sigma_{cl}$ or $R > 2 R_{200}$; blue dots) are distinguished. Galaxies without spectroscopic redshifts are denoted by grey dots.\label{gcmd}}
\end{figure*}

Using the CFHT MegaCam in the \emph{LSB mode}, we took 7 dithered exposures for each cluster field with total exposure time of 2940 seconds in the $g$ and $r$ bands, respectively. The pre-processing of the MegaCam data including overscan correction, bias subtraction and flat-fielding was done using the Elixir pipeline by the CFHT team \citep{mag04}. We removed fringe patterns from the processed images, and then we generated the dark sky image for each cluster which is a median combine of many exposures of the night. The scattered light from the primary mirror is removed by subtracting the dark sky image from object frames. The WCS information of each CCD was updated through the SCAMP program using the stars in the SDSS as a reference \citep{ber06}. The dithered images were resampled to have the same pixel scale ($0.185''$) and co-added into a deep image with the SWarp program \citep{ber02}.
The instrumental magnitudes were measured using the Source Extractor \citep[SExtractor;][]{ber96} and calibrated applying the calibration terms from the Elixir pipeline. The magnitudes (and surface brightnesses) were corrected for the foreground extinction based on \citet{sch11}. More details about the observation and data reduction of the KYDISC clusters will be described in Oh et al.~(in preparation).
After the image processing and stacking, the stellar full width at half maximum (FWHM) is about $0.8''$ and the field of view for each stacked image is about $87'\times 81'$ as displayed in Figure~\ref{targets}.

From the redshift information retrieved from the SDSS, NASA Extragalactic Database\footnote{http://ned.ipac.caltech.edu} and SIMBAD Astronomical Database\footnote{http://simbad.u-strasbg.fr/simbad/}, we selected cluster members by using the difference in recession velocity from the host cluster ($\Delta v_{rec}$) and the projected distance from the cluster center ($R$). We consider that the galaxies with  $|\Delta v_{rec}| \le 3 \sigma_{cl}$ and $R\le 2 R_{200}$ as cluster members, where $\sigma_{cl}$ and $R_{200}$ are the velocity dispersion and the virial radius of a given cluster, respectively. The $\sigma_{cl}$ and $R_{200}$ values  for each cluster are listed in Table~\ref{clinfo}. 

\begin{figure}[!ht]
\centering
\plotone{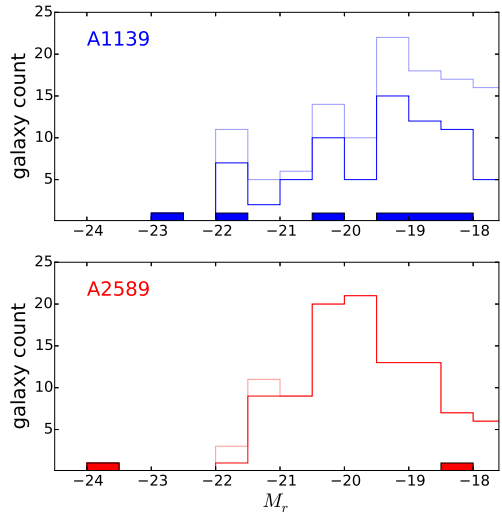}
\caption{Galaxy luminosity functions of the target clusters. Only the galaxies with $|\Delta v_{rec}| \le 3 \sigma_{cl}$ are used. Different histogram types represent different clustercentric distance cuts: light open, dark open, and dark filled histograms for $R\le 2 R_{200}$, $R\le R_{200}$, and $R\le 50$ kpc, respectively.\label{glf}}
\end{figure}

\begin{figure*}[!ht]
\centering
\includegraphics[width=0.95\textwidth]{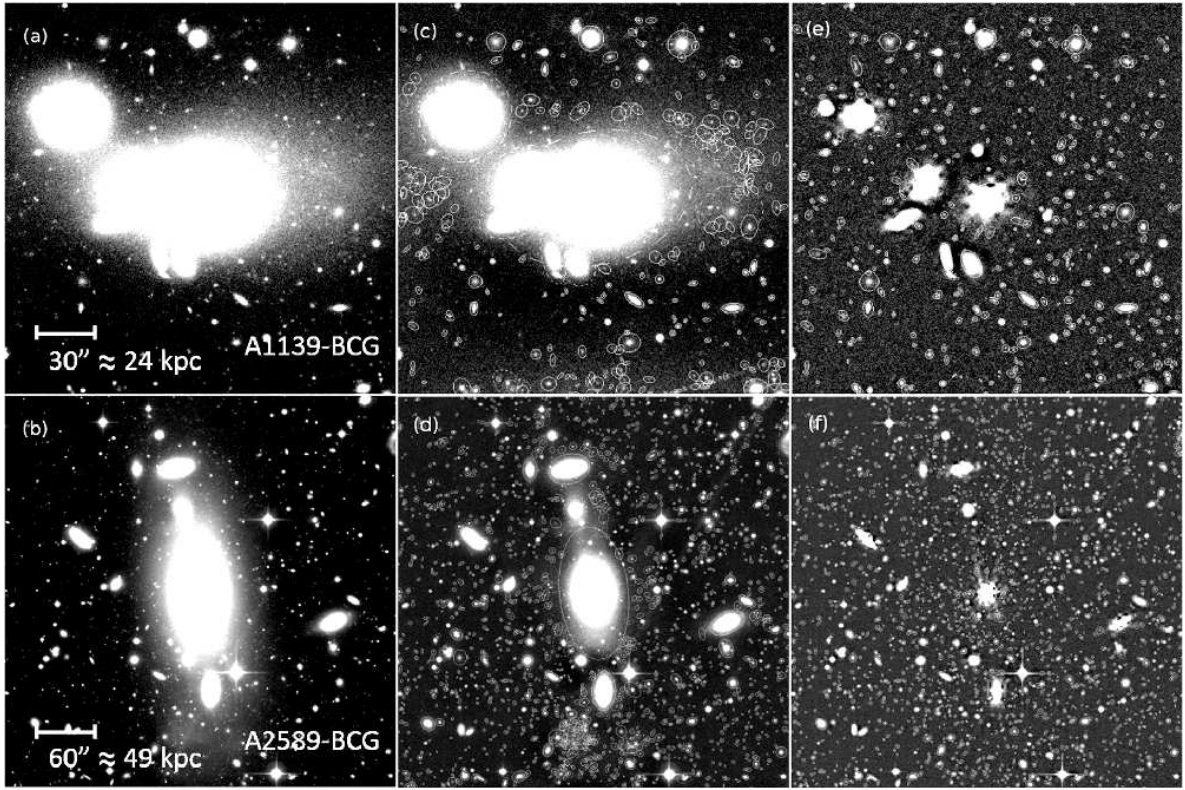}
\caption{The process to mask contaminating light from companion galaxies and background objects.(a) and (b): The original images centered on the BCGs. (c) and (d): Mask apertures of the contaminating objects using big background-meshes. (e) and (f): Mask apertures of the contaminating objects using small background-meshes. The upper panels are for the {\bcga}, while the lower panels are for the {\bcgb}. The apertures for the BCGs are also drawn, but they are not used for the masking.\label{masking}}
\end{figure*}

Figure~\ref{gcmd} shows the galaxy color-magnitude diagrams of the target clusters, and their galaxy luminosity functions are compared in Figure~\ref{glf}. The $r$-band magnitude gap between the two brightest galaxies in each cluster ($m_{12}$) is large in A2589 ($m_{12}=1.81$), while it is relatively small in A1139 ($m_{12}=0.60$). A2589 is not a fossil cluster according to the definition of \citet[][should be $m_{12}\ge2.0$]{jon03}, but it seems to be \emph{fossil-like}. If the idea is true that fossil clusters have been assembled very long time ago and grow only through minor mergers at $z<1$ \citep{don04,don05,kho06,von08,dar10}, the large $m_{12}$ value is consistent with the previous studies arguing that A2589 is relaxed well.
In Figure~\ref{glf}, A1139 has 6 galaxies brighter than $M_r = -18$ within 50 kpc from the cluster center, while A2589 has only 2 galaxies (including the BCG) in the same condition.
From the very high density of the A1139 center, one may speculate that the bright galaxies in the center of A1139 will merge into a single massive BCG in the future.

In this paper, we study  the BCGs of the two target clusters and also investigate the bright ($M_r \le -21$) galaxies at similar redshifts ($\Delta z \le 0.0067$) for comparison. The number of the comparison galaxies is 19 in the A1139 field and 21 in the A2589 field.  Among the comparison galaxies, the number of early-type galaxies is 9 and 8 in the A1139 and A2589 fields respectively, in our visual inspection.

\begin{figure*}[!ht]
\centering
\includegraphics[width=0.3\textwidth]{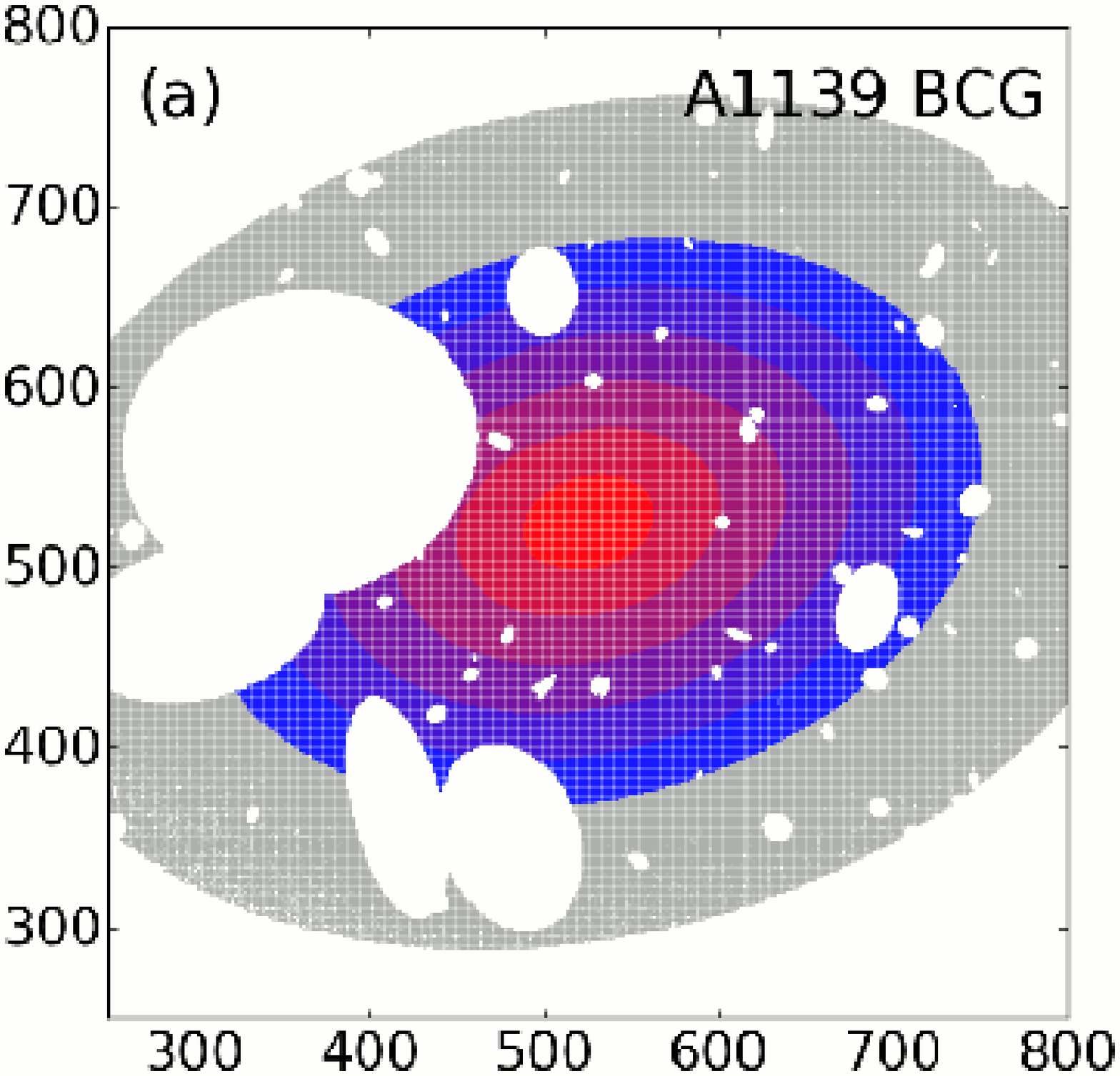}
\includegraphics[width=0.3\textwidth]{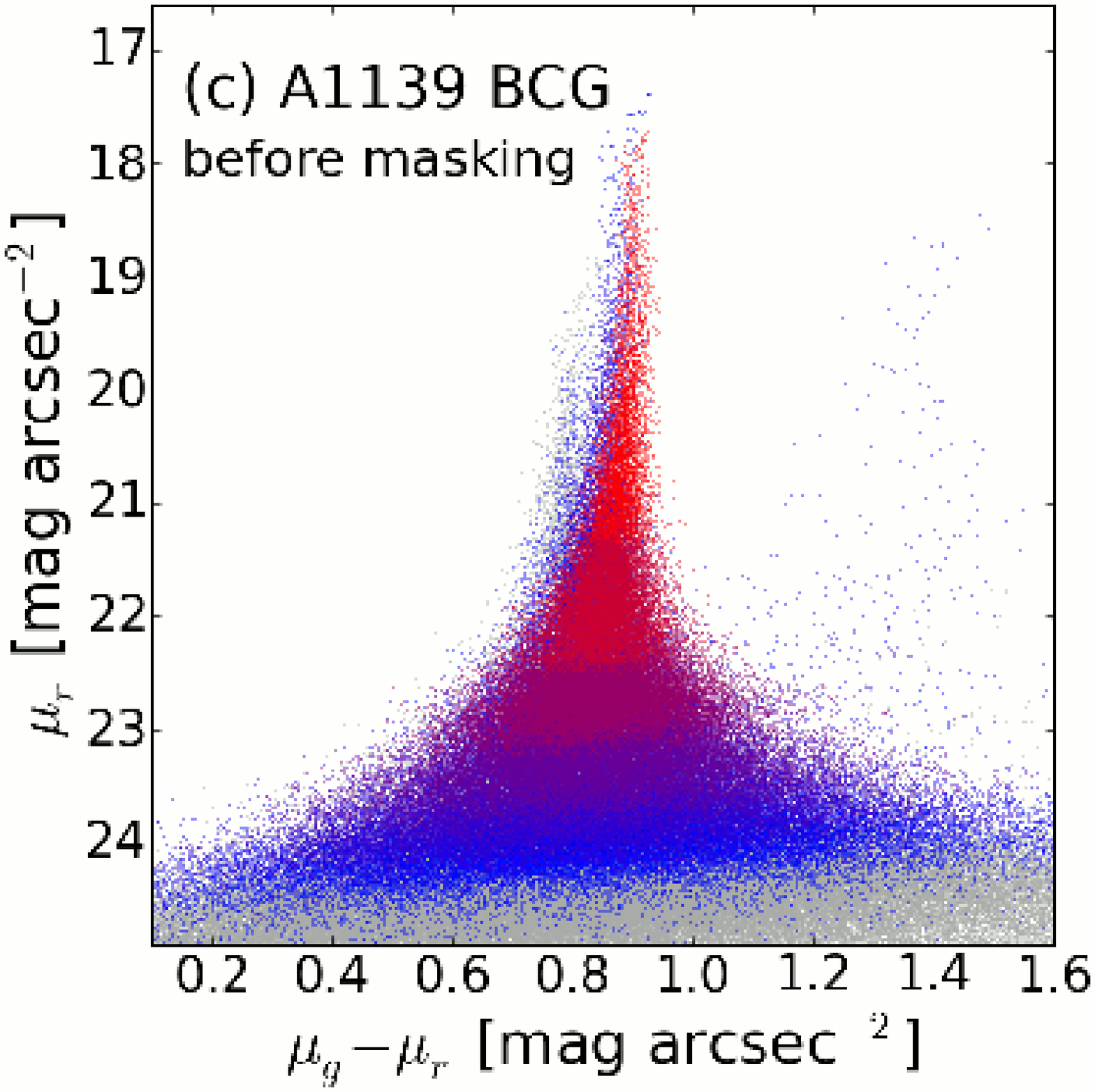}
\includegraphics[width=0.3\textwidth]{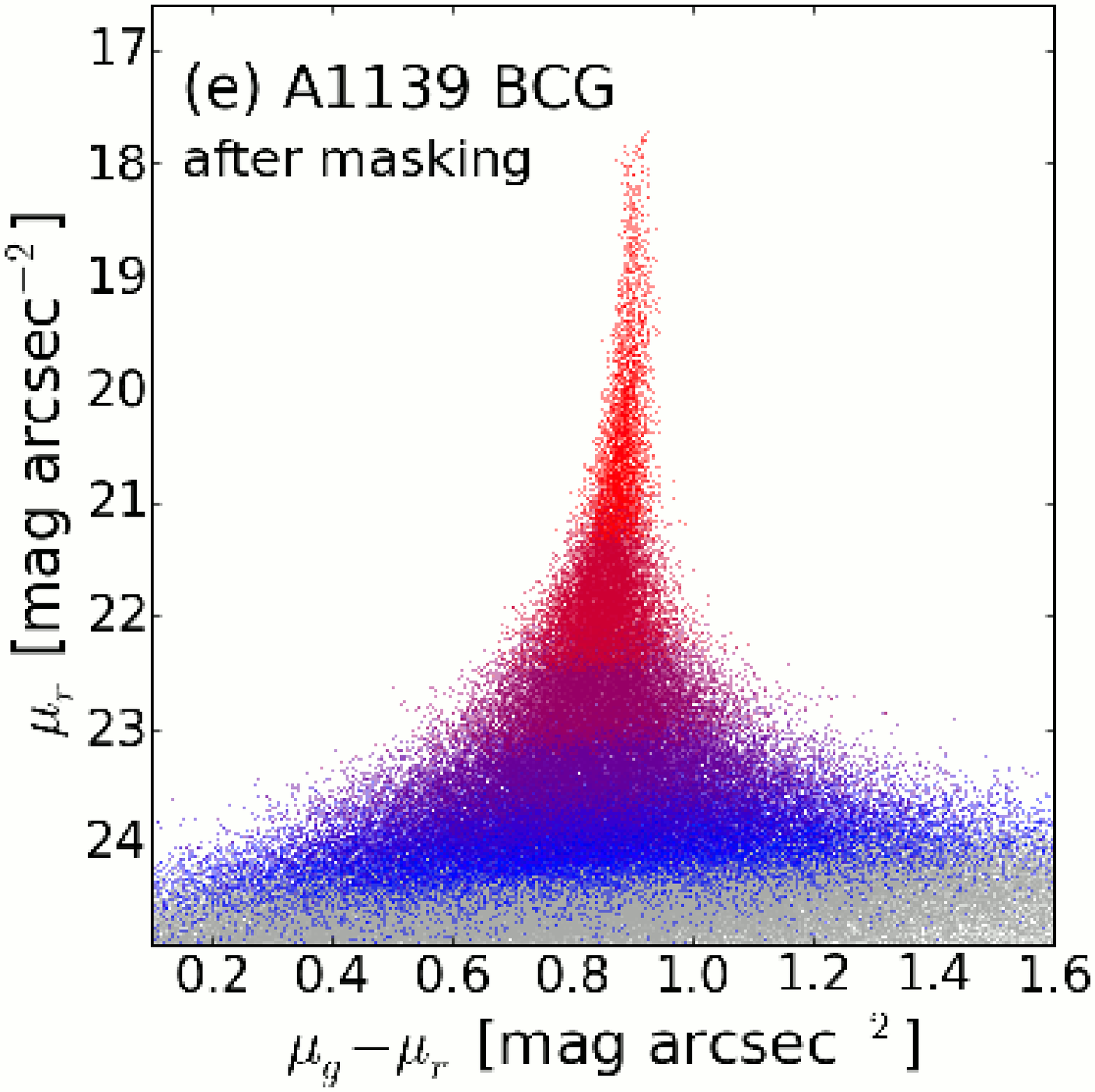}
\includegraphics[width=0.3\textwidth]{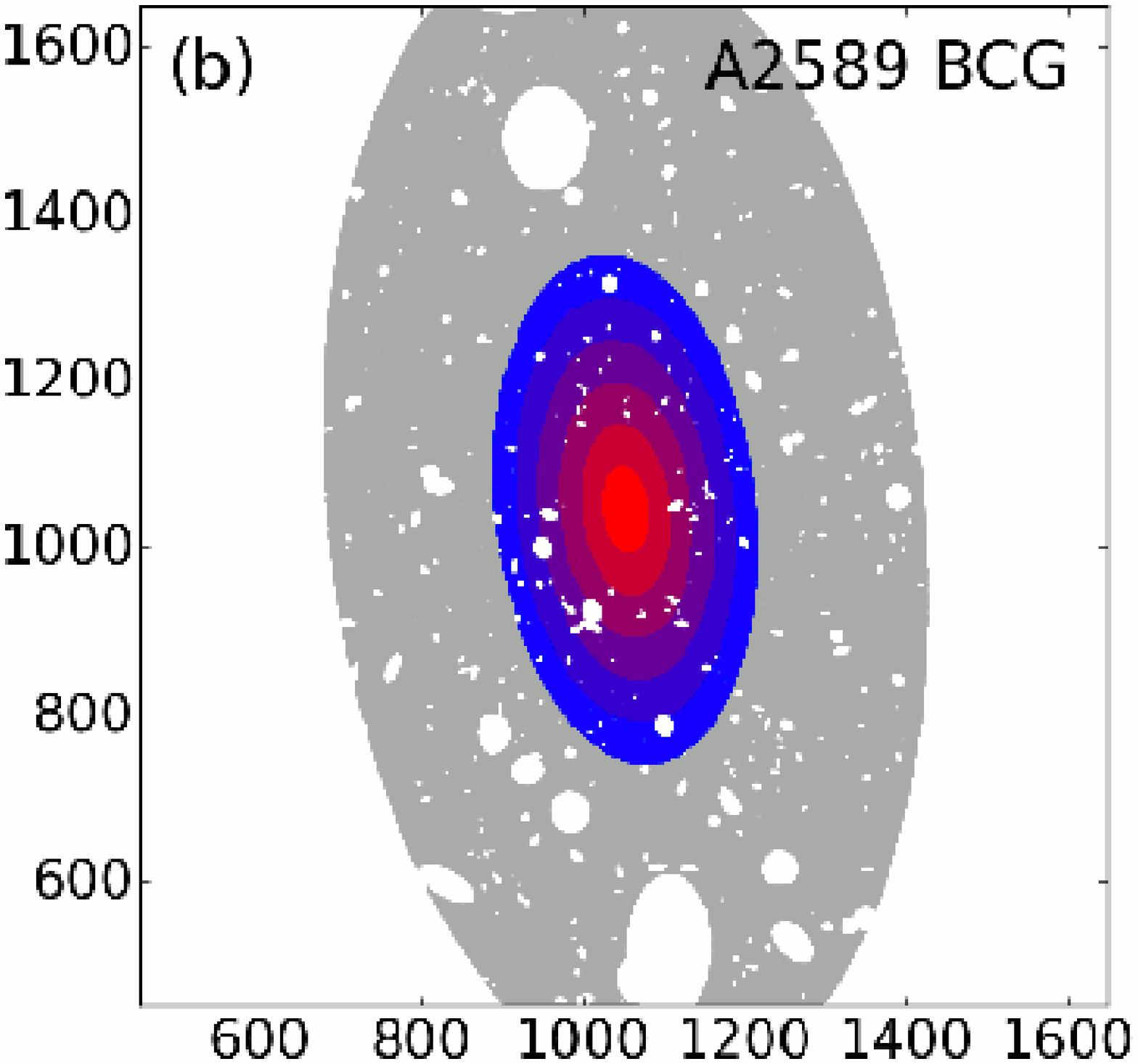}
\includegraphics[width=0.3\textwidth]{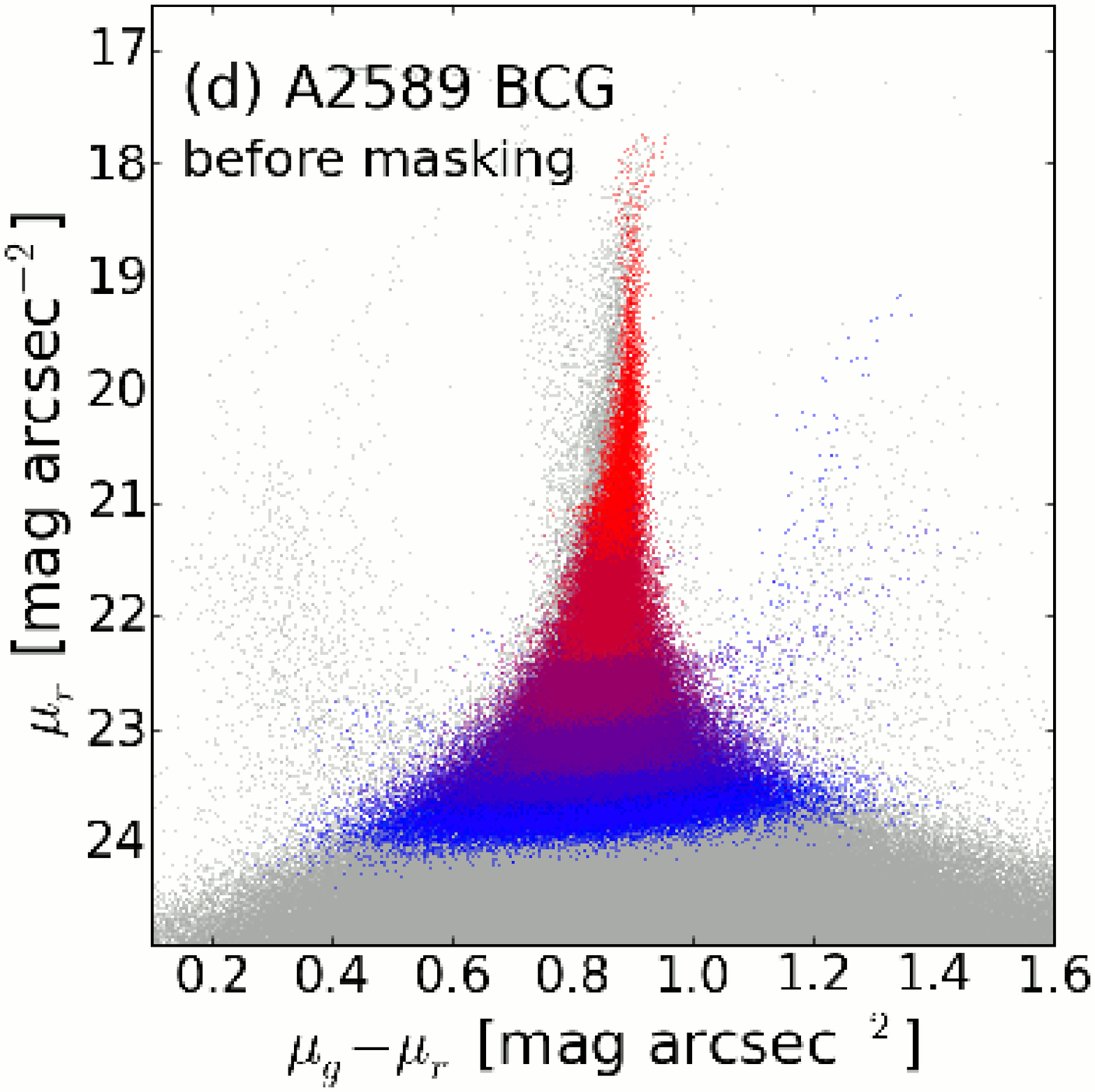}
\includegraphics[width=0.3\textwidth]{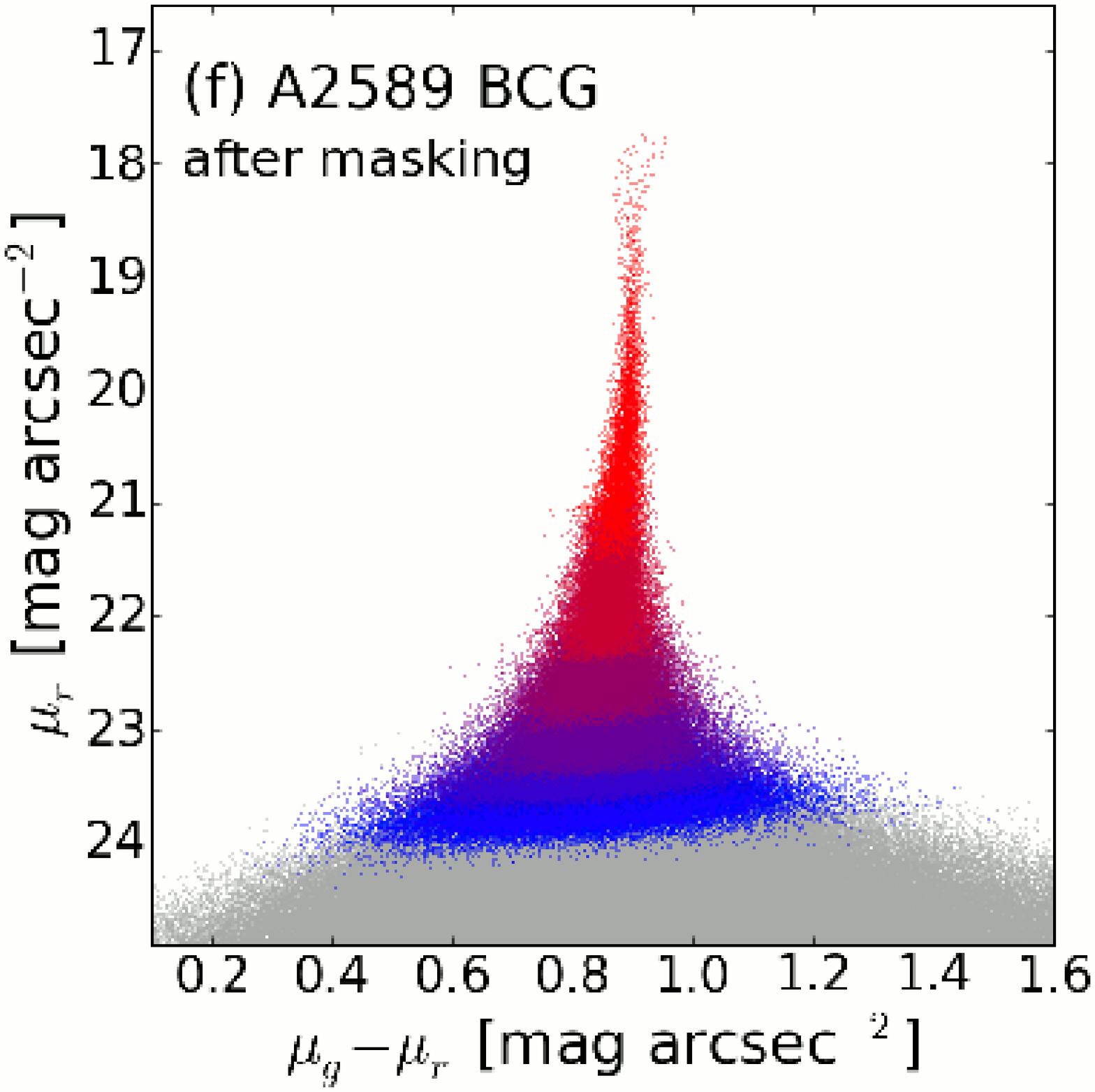}
\caption{(a) and (b): Pixel maps of the target BCGs after the masking. (c) and (d): Pixel color-magnitude diagrams (pCMDs) of the target BCGs before the masking. (e) and (f): pCMDs of the target BCGs after the masking. The colors of the dots are coded by the spatial distribution of the pixels, based on the isophotal ellipses of each BCG.\label{pmap}}
\end{figure*}

\begin{figure*}[!ht]
\centering
\includegraphics[width=0.45\textwidth]{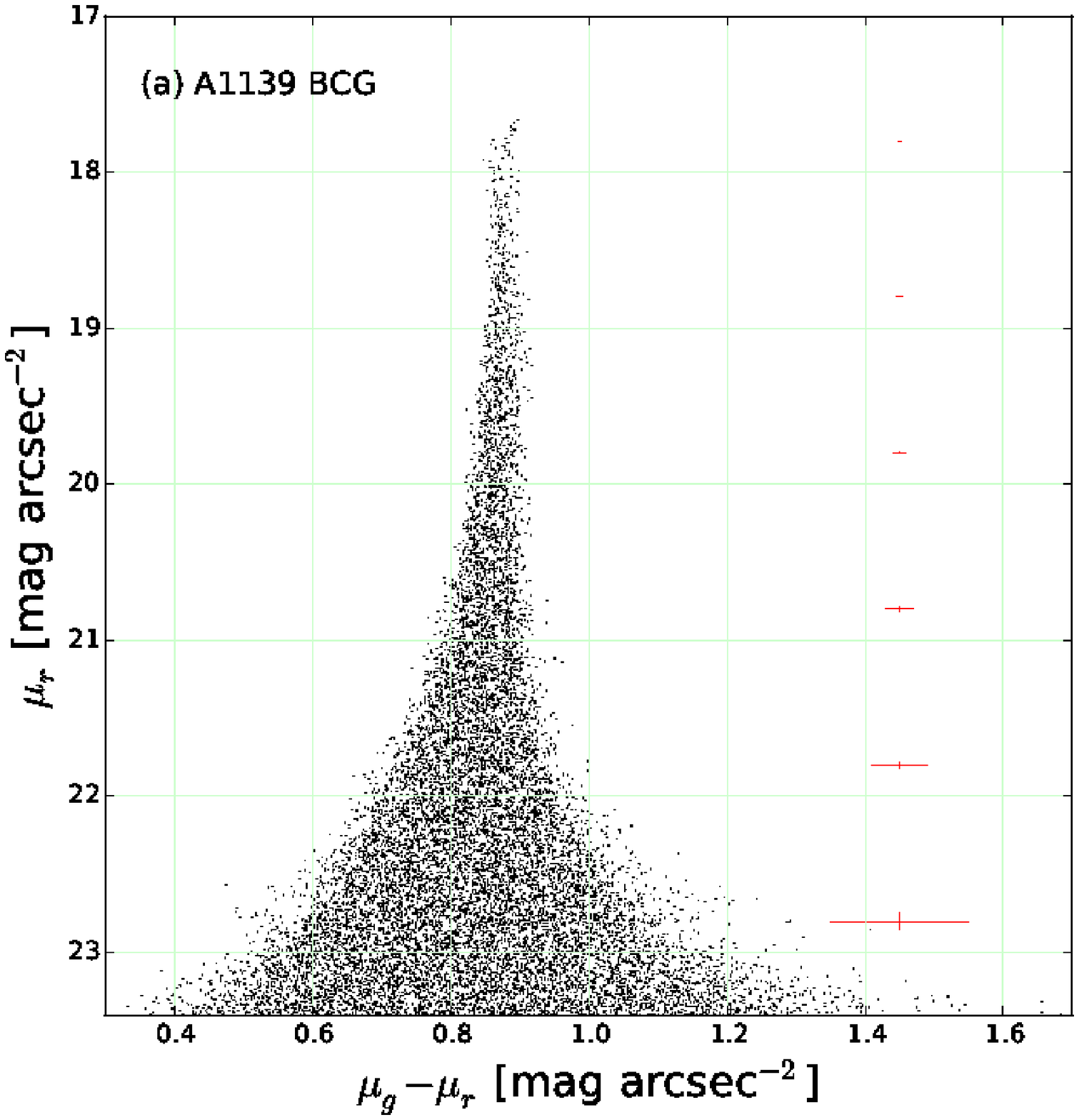}
\includegraphics[width=0.45\textwidth]{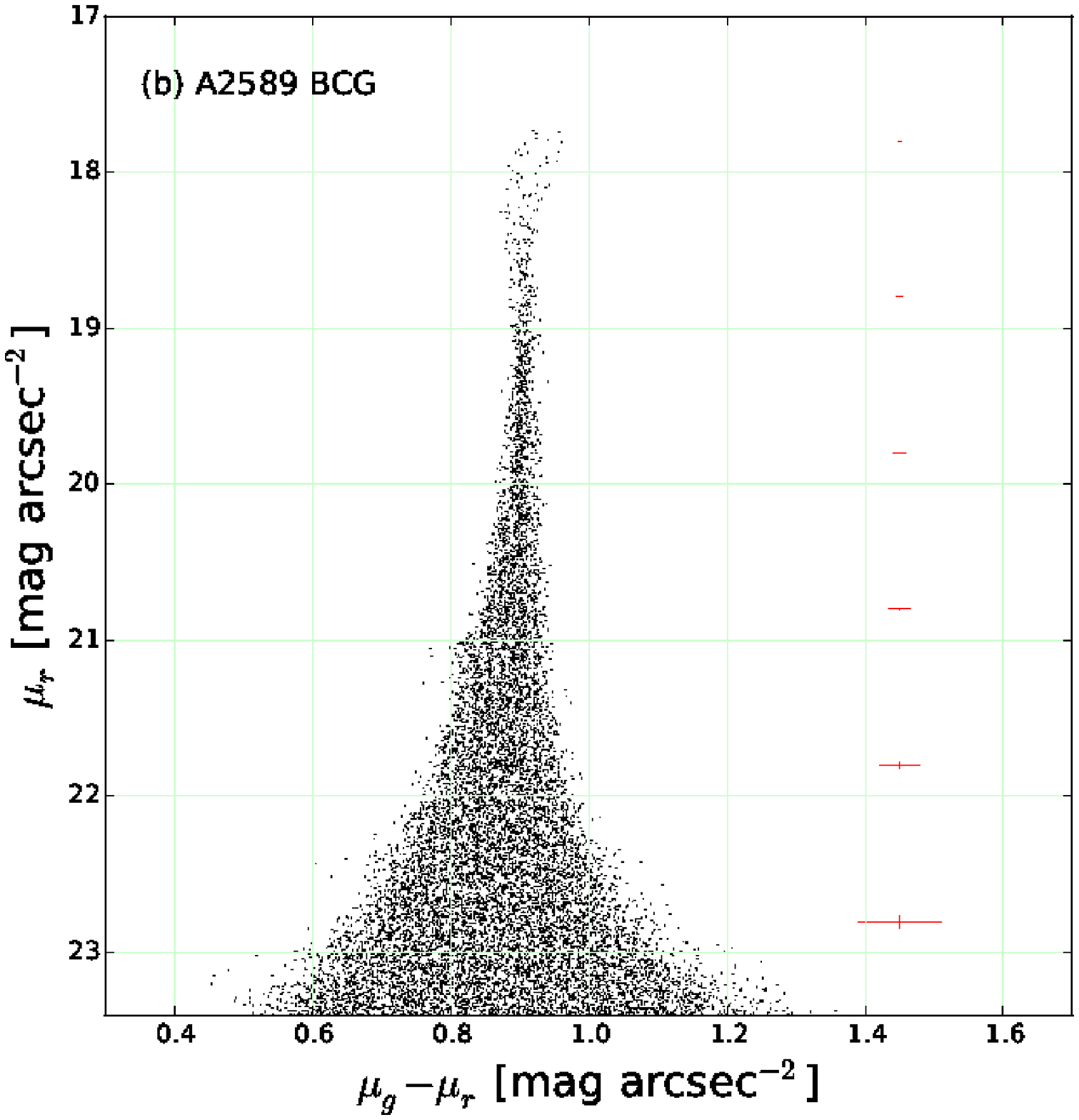}
\includegraphics[width=0.45\textwidth]{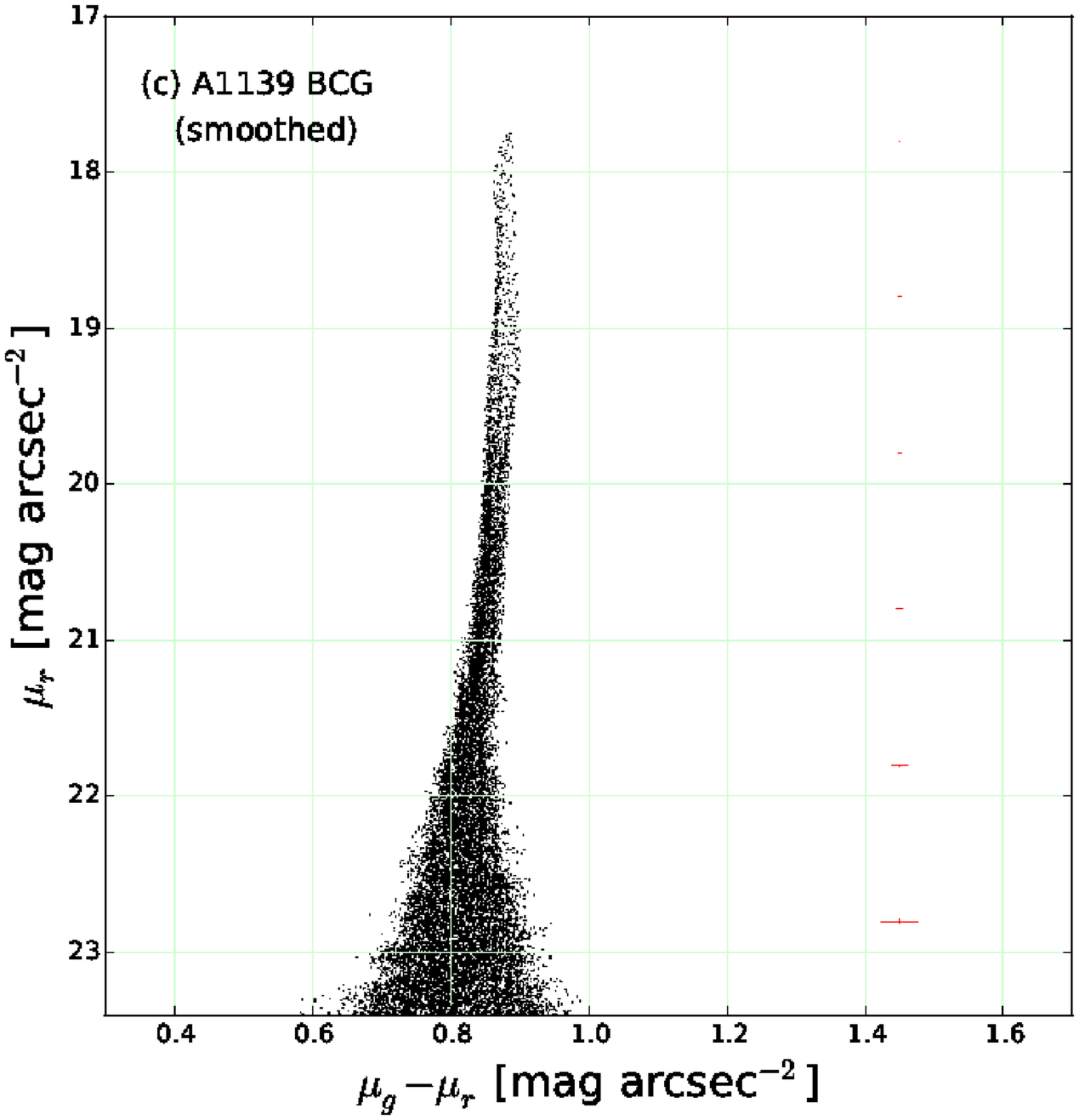}
\includegraphics[width=0.45\textwidth]{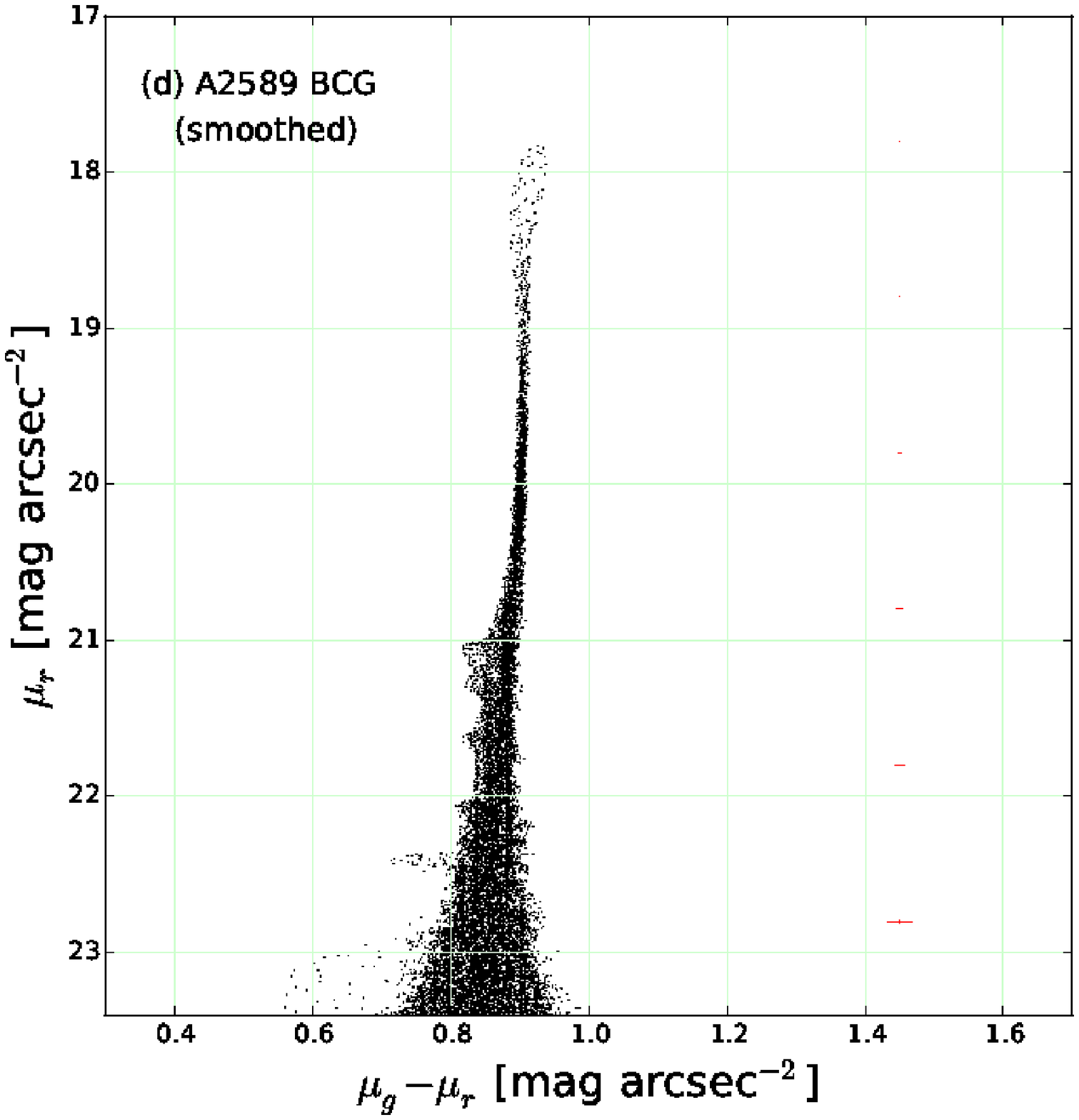}
\caption{The pCMDs of the BCGs before (upper panels) and after (lower panels) the pixel smoothing with an aperture of $0.8''$. Representative photometric errors in a single pixel at several $\mu_r$ levels are denoted.\label{pcmd}}
\end{figure*}

\begin{figure*}[!ht]
\centering
\includegraphics[width=0.95\textwidth]{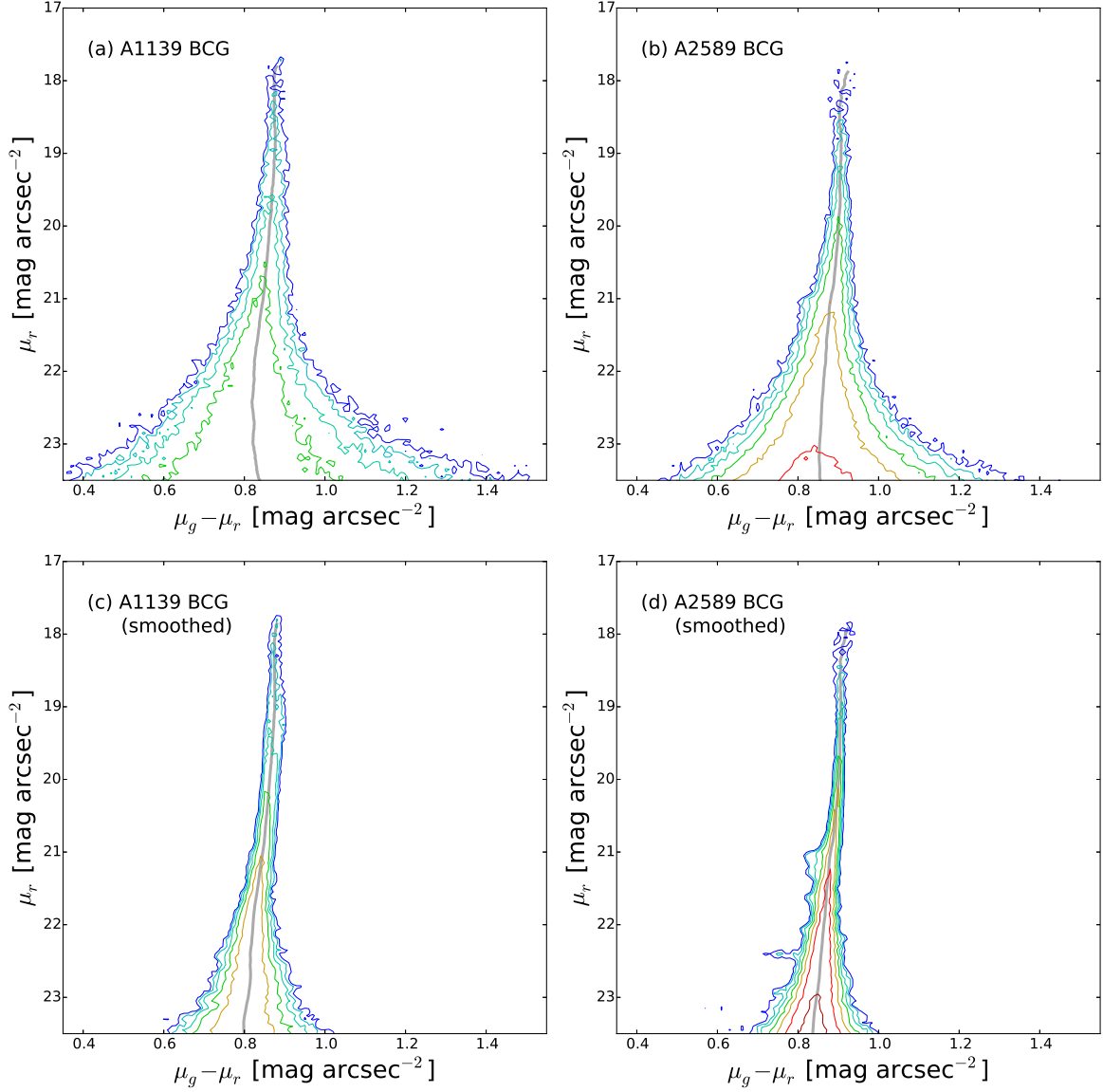}
\caption{The contour map version of Figure~\ref{pcmd}. The contours follow the log-scale density of pixels in each pCMD. The grey line connects the mean pixel color at given $\mu_r$ in each panel.\label{pcmdcon}}
\end{figure*}

\begin{figure*}[t]
\centering
\includegraphics[width=0.41\textwidth]{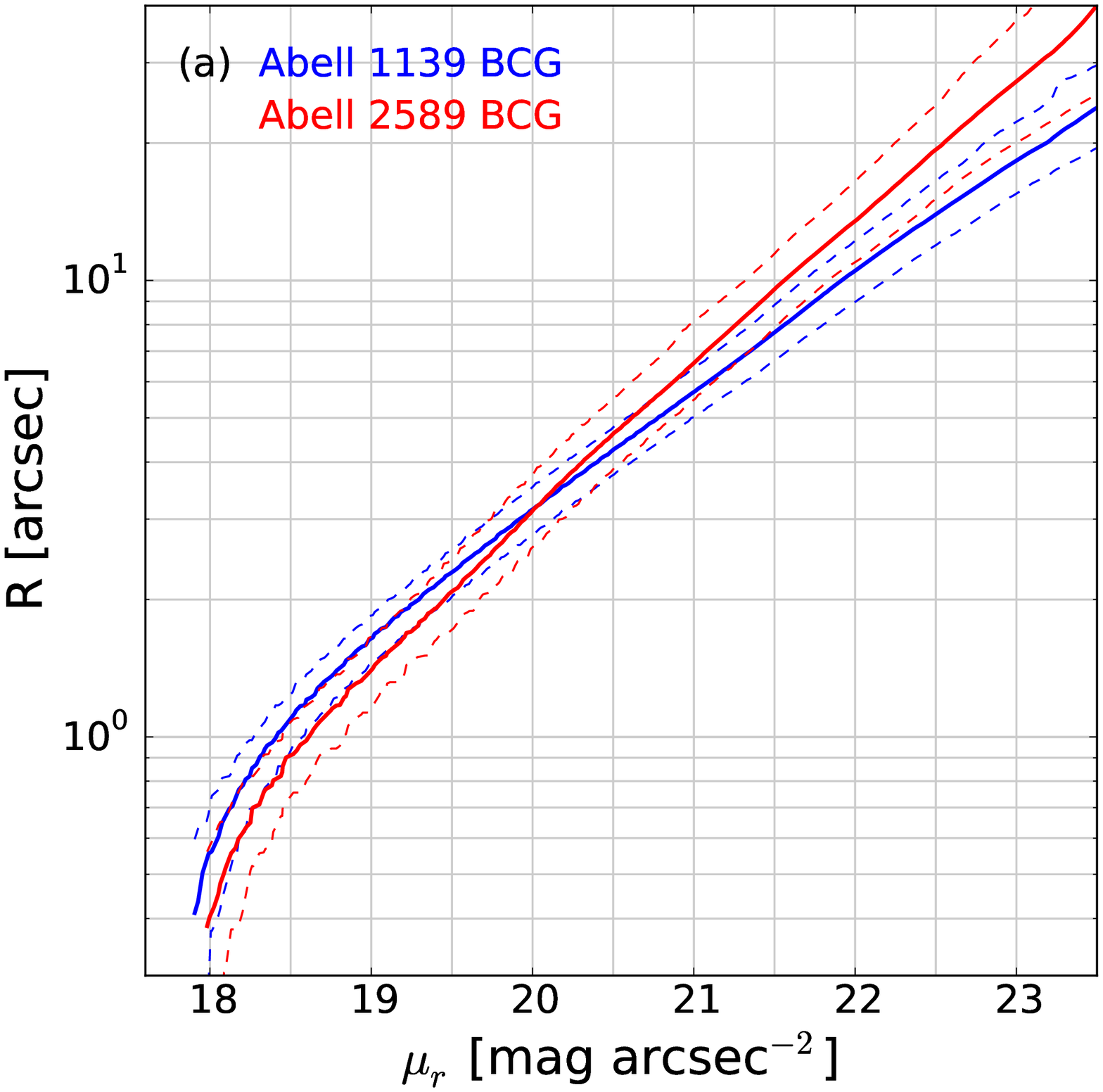}
\includegraphics[width=0.41\textwidth]{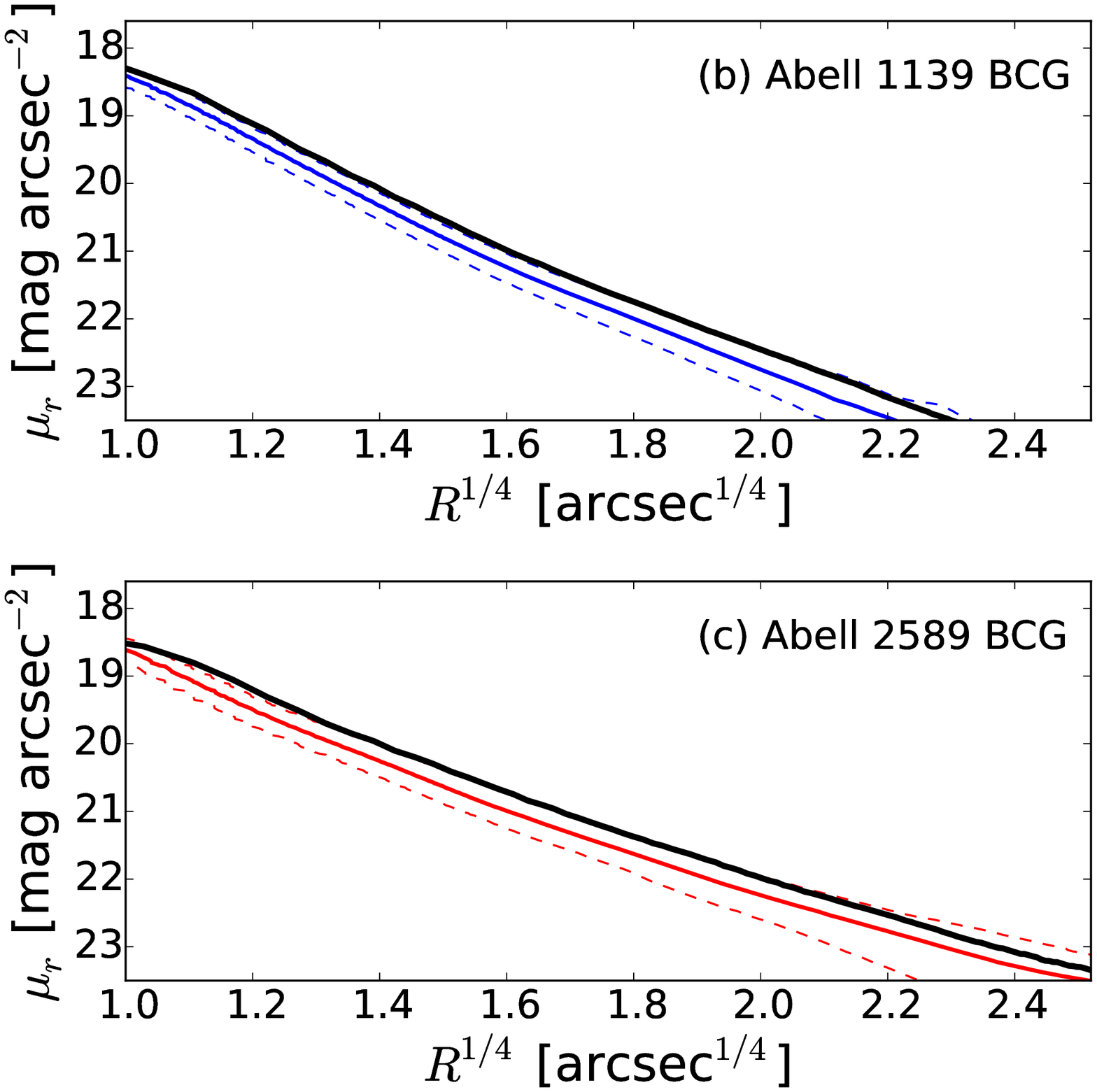}
\caption{(a) Correlations between pixel surface brightness ($\mu_r$) and distance to the BCG center (R) in the {\bcga} (blue lines) and the {\bcgb} (red lines). The solid lines are the median distances, while the dashed lines show the minimum ($1\%$) and maximum ($99\%$) distances. (b) and (c): Comparison with the surface profiles from the isophotal ellipses fitting (thick black lines). The blue and red lines are the same as the lines in (a).\label{radmu}}
\end{figure*}

\section{ANALYSIS}

\begin{figure}[t]
\centering
\plotone{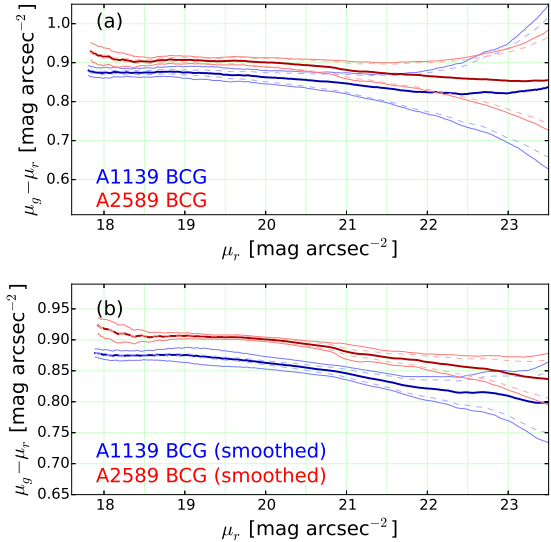}
\caption{(a) Backbones of the pCMDs. The thick solid line connects the mean $g-r$ colors at given $\mu_r$ for each BCG's pCMD. The thin solid lines show the standard deviation of pixel colors, while the dashed lines indicate the typical photometric error of a single pixel. (b) The pCMD backbones after the pixel smoothing.\label{backbone}}
\end{figure}

\begin{figure}[!t]
\centering
\plotone{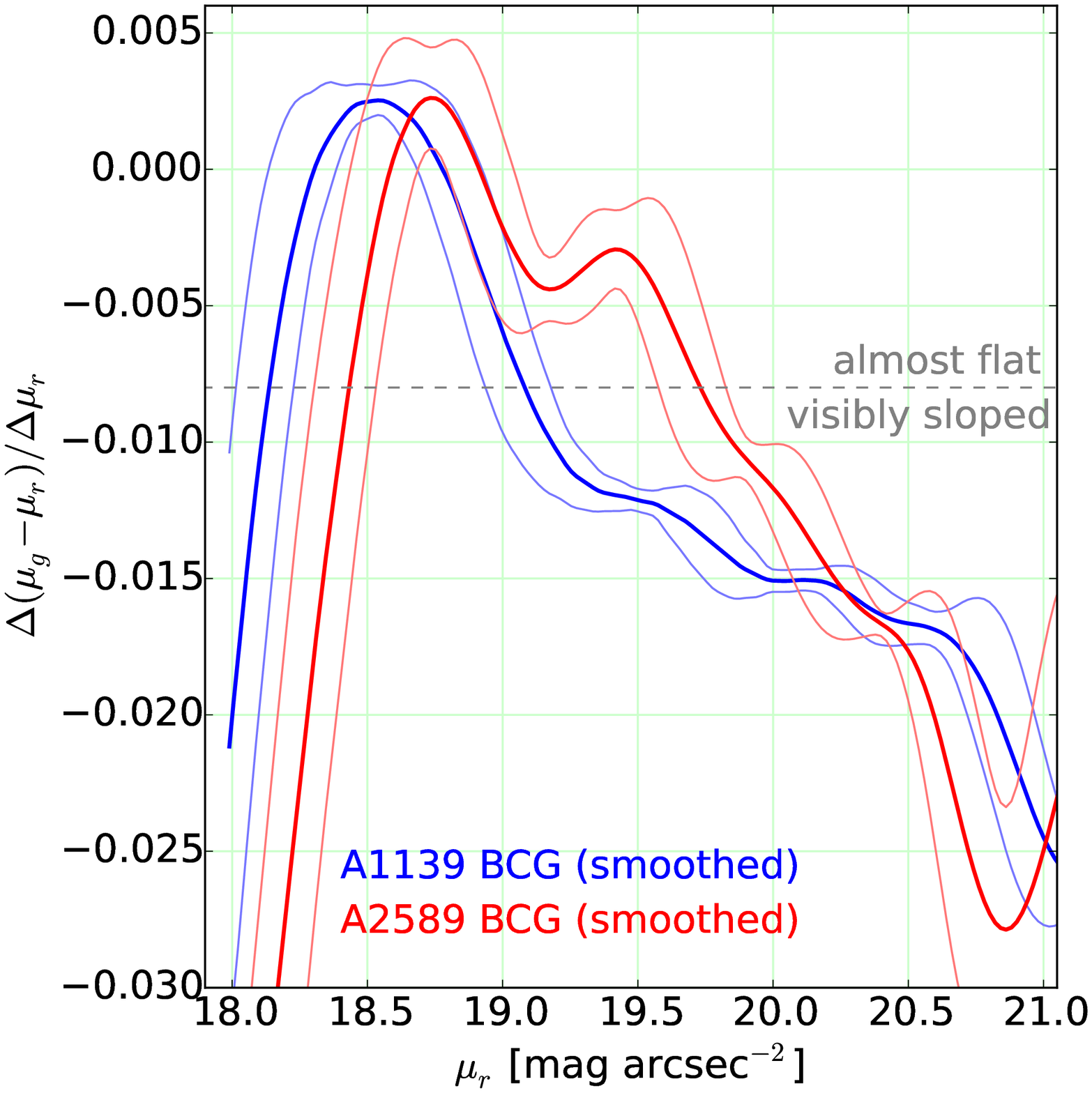}
\caption{Slope variation of the pCMD backbones. The thick lines indicate the mean slope within the smoothing range ($\pm0.2$ {\umu}), while the thin lines show the slope deviation within the smoothing range. These are based on the smoothed pCMDs.\label{slope}}
\end{figure}

\subsection{Pixel Analysis Procedure}\label{anal}

Since the target BCGs have many companion galaxies as well as background galaxies, the light from those contaminating objects must be removed before the pixel analysis.
The process to mask contaminating light is as follows:
\begin{enumerate}
 \item Trim a sufficiently large area around the BCG: $194'\times194'$ for the {\bcga} and $379'\times379'$ for the {\bcgb}.
 \item Extract objects using the SExtractor. Detect bright objects including the BCG by adopting background meshes with sufficiently large sizes (larger than the effective radius of the BCG). 
 \item Detect small objects by adopting background meshes with small sizes (moderately larger than the typical seeing size). This enables to detect faint and small objects within the extent of the BCG, because the diffuse light of the BCG is regarded as background in this setting.
 \item Mask the pixels within the apertures of any detected non-BCG objects.
\end{enumerate}
Figure~\ref{masking} shows the images for the steps 1 -- 3.

The pixel maps after the masking process and the pCMDs before and after the masking are presented in Figure~\ref{pmap}. The pCMDs before the masking show many outlying dots because of the bright pixels far from the BCG center (i.e., the light from companion or background galaxies). After the masking, those outlying dots are mostly removed, and clean pCMDs of the main bodies of the BCGs are left.

Since the typical stellar FWHM is $0.8''$ while the pixel scale is $0.185''$, the light in each pixel is significantly blurred by the seeing effect. Thus, it is efficient to smooth the pixels with an aperture of the seeing size ($0.8''$), because it improves the S/N in each pixel with little concern of decreasing resolution.
Figure~\ref{pcmd} compares the pCMDs before and after the pixel smoothing and Figure~\ref{pcmdcon} shows the density contour map version of the pCMDs. The scatters of the pCMDs significantly decrease after the pixel smoothing, which makes the pCMDs look tighter. Moreover, we find several fine features in the smoothed pCMDs, which are hardly distinguished before the pixel smoothing. That is, the pixel smoothing reveals the fine features in the pCMDs more clearly by improving the S/N in individual pixels. We limit our analysis to the pixels brighter than $\mu_r=23.5$ mag arcsec$^{-2}$ hereafter, at which the typical photometric uncertainty of a single pixel is $0.02 - 0.03$ mag arcsec$^{-2}$ after the smoothing.

Figure~\ref{radmu} shows the correlations between pixel surface brightness in the $r$ band ($\mu_r$) and distance to the BCG center in the two BCGs. The minimum and maximum distances (lower and upper dashed lines) approximately correspond to the minor and major axis lengths of the ellipse with given $\mu_r$.
In the pCMD analysis, $\mu_r$ is a key parameter, along which most quantities are compared. However, when we need to compare quantities as a function of radius rather than $\mu_r$, the conversion will be done using the correlations in Figure~\ref{radmu}(a). While $\mu_r$ itself represents the stellar luminosity density at a given pixel, this conversion is often useful to understand the physical implications of the results in the viewpoint of galaxy structure.
The panels (b) and (c) in  Figure~\ref{radmu} compare the $\mu_r$ -- $R$ correlations with the surface brightness profiles derived from the classical isophotal ellipses fitting (with fixed center and position angle, but varying ellipticity). As expected, the maximum distance lines mostly agree with the ellipses fitting profiles, although the discrepancy increases as $\mu_r$ gets fainter. Note that the ellipses fitting profiles depend on the parameter setup and fitting options to some extent, while the the $\mu_r$ -- $R$ correlations can not be changed because they are non-parametrically estimated.

\subsection{Advantages over Classical Methodology}

Here, we discuss several advantages of the pCMD analysis over the classical structure analysis based on the ellipses fitting.
The first and most important strength is that the pCMD analysis can be consistently applied to galaxies with any kind of morphology. The pCMD analysis is non-parametric: it does not depend on any function-based fitting and thus it can be equally used not only for single-component galaxies (elliptical galaxies and pure-disk galaxies) but also for complicated galaxies with multiple components (lenticular, spiral, cD and even irregular galaxies and mergers). Although the 2D-structure analysis using multiple components fitting is widely conducted today and it is worthy in understanding galaxy structures, the pCMD analysis is an approach complementary to such classical methods, which is simpler, more efficient, and less dependent on pre-assumptions about the structural components in target galaxies. The pCMD analysis is particularly efficient in detecting fine sub-structures in a galaxy, as presented in Section~\ref{finefeat}.

Second, the pCMD analysis is based on the $\mu$ versus color parameter space, but is also extended to spatial structure analysis. In other words, it considers  the photometric and structural properties at the same time, from which great variety of features appear according to galaxy morphology \citep{lan07}.
The pCMD method is not a simple profile analysis, although it may look very similar to a 1D profile analysis when it is used for normal elliptical galaxies. The more complicated the morphology of a target galaxy is, the more distinct results the pCMD analysis returns.
This means adversely that the pCMD analysis may be less worthy for normal elliptical galaxies, which can be almost perfectly replaced with surface profile analysis along semi-major axis. However, our target galaxies are BCGs, which are known to be somewhat different from normal elliptical galaxies and more complicated in their structures \citep[e.g.,][]{von07}.

\begin{deluxetable}{cccc}
\tablenum{2} \tablecolumns{4} \tablecaption{Slope Information of the pCMD Backbones after Plateaus} \tablewidth{0pt}
\tablehead{ & & {\bcga} & {\bcgb} }
\startdata
&  $\mu_r$ & 19.1 {\umu} & 19.7 {\umu} \\
& $g-r$ & 0.874 & 0.904 \\
At the end of plateau & $R$ & $1.7''$ ($\approx 1.3$ kpc) & $2.4''$ ($\approx 2.0$ kpc) \\
& age $^{(a)}$ & 9.41 Gyr & 11.74 Gyr \\
& Z  $^{(b)}$ & 0.031 & 0.038 \\
\hline
&  $\mu_r$ & 22.1 {\umu} & 22.7 {\umu} \\
& $g-r$ & 0.820 & 0.854 \\
At the end of plateau & $R$ & $10.9''$ ($\approx 8.6$ kpc) & $22.2''$ ($\approx 18.2$ kpc) \\
+ 3.0 {\umu} & age & 6.52 Gyr & 8.18 Gyr \\
& Z & 0.019 & 0.026 \\
\hline
& $\Delta(g-r)$ & $-0.054$ & $-0.050$ \\
For $\Delta\mu_r=$ & $R_{outer}/R_{inner}$ & 6.4 & 9.3 \\
3.0 {\umu} & $\Delta age$ & $-2.89$ Gyr & $-3.56$ Gyr \\
& $\Delta$[M/H] & $-0.21$ & $-0.16$ \\
\enddata
\tablecomments{(a) At fixed metallicity of Z = 0.04. (b) At fixed age of  12 Gyr. All values are based on the smoothed pCMDs.}
\label{slopeinfo}
\end{deluxetable}

Third, in case of using the pCMD analysis as a replacement of the classical surface (brightness and color) profile analysis, one of the major differences between the pCMD analysis and the classical analysis is that the former is mainly conducted along surface brightness ($\mu$), while the latter is mostly based on radius ($R$; or semi-major axis). The information of physical scale is very important to understand galaxy structures and thus it should be always considered and compared even while we conduct the pCMD analysis.
However, when we compare between galaxies with different masses and sizes, the normalized radius ($R/R_e$; where $R_e$ is the effective radius) will be a better standard than the simple $R$.
By comparing the surface profiles along $R/R_e$, we can check the homology of the target galaxies, which may be connected with the similarity in their formation processes.
If the target galaxies are homologous (that is, formed in similar processes but different scales), they will show very similar surface profiles along $R/R_e$, although their profiles along $R$ may differ from each other due to their different scales.

However, in case of BCGs, it is somewhat difficult to accurately estimate their effective radii, because there are typically so many contamination sources in the center of a cluster, such as many bright satellite galaxies and intracluster light \citep[e.g.,][]{bur15}. Moreover, the dependence of the isophotal ellipses fitting on the parameter setup and fitting options should be more carefully considered for BCGs, not only because there are many contamination sources but also because the internal structure itself of a BCG tends to be more complicated than a normal elliptical galaxy. In this case, $\mu$ is a good replacement of $R/R_e$ for BCGs, which enables us to conduct a non-parametric structure analysis and thus is useful to check the homology of the target BCGs. In a single component light profile, $\mu$ can be a fine proxy of $R/R_e$, as shown in Figure~\ref{radmu}.

In summary, the pCMD analysis is a simple, efficient, and non-parametric tool to simultaneously investigate the photometric and structural properties of target galaxies with any morphological types. 
It is not true that the pCMD analysis is superior to the classical structure analysis method in every aspect, because the latter obviously has its own strengths; for example, the 2D multi-components fitting method is better for clear structure decomposition. The two approaches are complementary to each other, and the pCMD analysis gives us insight on the formation histories of target galaxies in a different viewpoint.

\section{RESULTS}\label{result}

\subsection{Backbones of the pCMDs}\label{backb}

For quantitative comparison of the pCMDs, we estimate the mean $g-r$ colors of pixels as a function of $\mu_r$, which we call the \emph{backbones} of the pCMDs.
In Figure~\ref{backbone}, the backbone of the {\bcga} appears to be bluer than that of the {\bcgb} at any $\mu_r$: the difference in mean color at given $\mu_r$ between the two BCGs is as large as $\Delta(g-r)\sim0.02-0.04$ when the pCMDs are smoothed.
It is noticeable that the total color of the {\bcga} ($g-r=0.842$) is similar to (or slightly redder than) that of the {\bcgb} ($g-r=0.838$) by $\Delta(g-r)\sim0.004$, unlike the trend at given $\mu_r$. This is because the {\bcga} has a smaller number fraction of faint pixels (typically having relatively blue colors) than the {\bcgb}, which is revealed in Figure~\ref{radmu} and will be also shown in Section~\ref{plfsec}.

As well as the mean colors of the backbones, their curvatures also differ from each other. We compare the slope variations of the pCMDs in Figure~\ref{slope}. We first calculated the backbone slope between $\mu_r-0.2$ {\umu} and $\mu_r+0.2$ {\umu} at given $\mu_r$, based on the smoothed pCMDs. Since the raw slope values show noisy fluctuations along $\mu_r$, we smoothed the slopes within the range of $\mu_r\pm0.2$ {\umu}. 
Overall, the backbones of the pCMDs show very shallow slopes at their bright parts (hereafter, \emph{plateaus}), except that the pCMD backbone of the {\bcgb} shows large negative slopes at the brightest end ($\mu_r<18.4$ {\umu}).
However, as $\mu_r$ increases (gets fainter), the {\bcga} backbone falls to the relatively blue domain earlier than the {\bcgb} backbone does. 

\begin{figure*}[!t]
\centering
\includegraphics[width=0.45\textwidth]{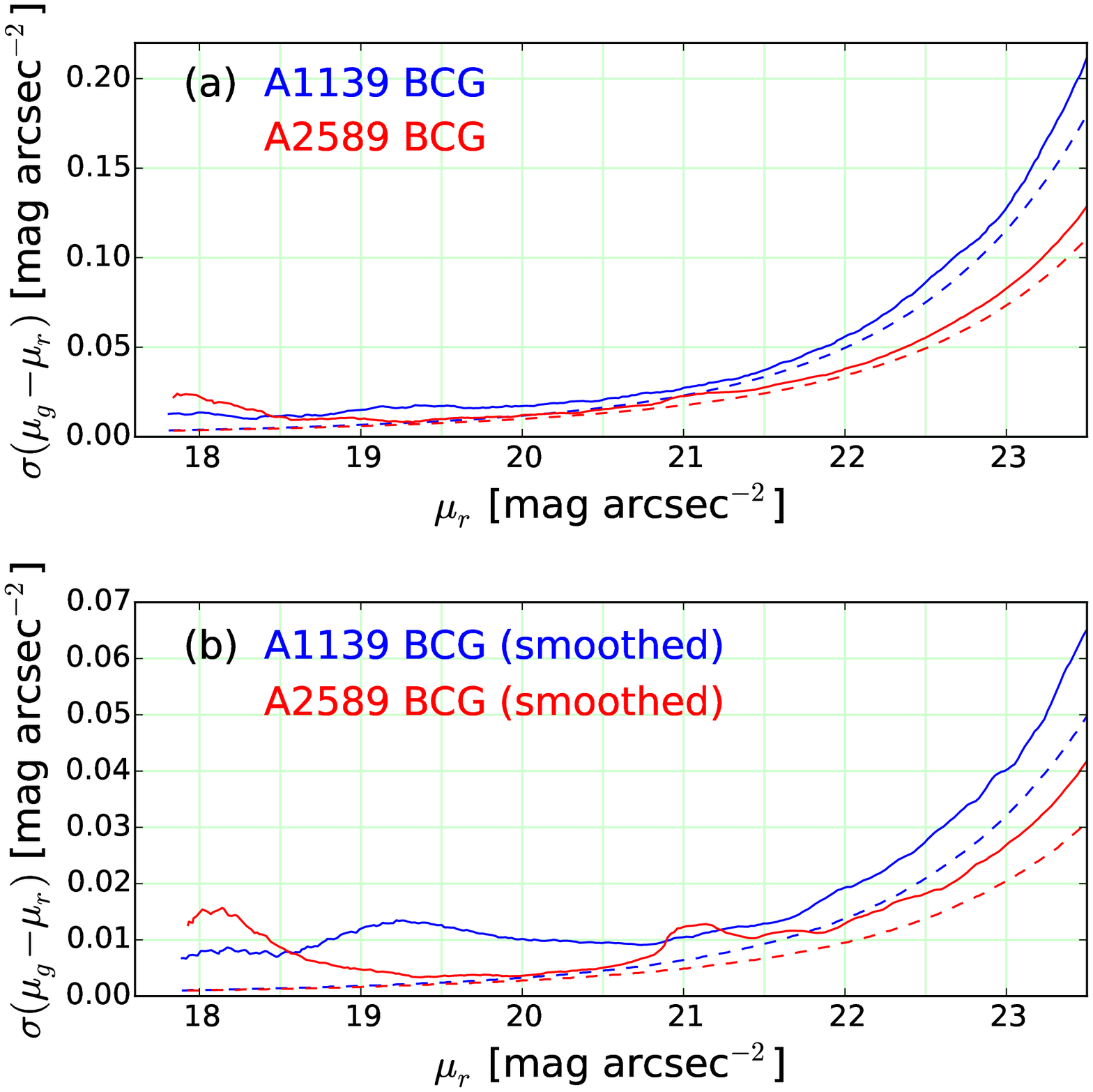}
\includegraphics[width=0.45\textwidth]{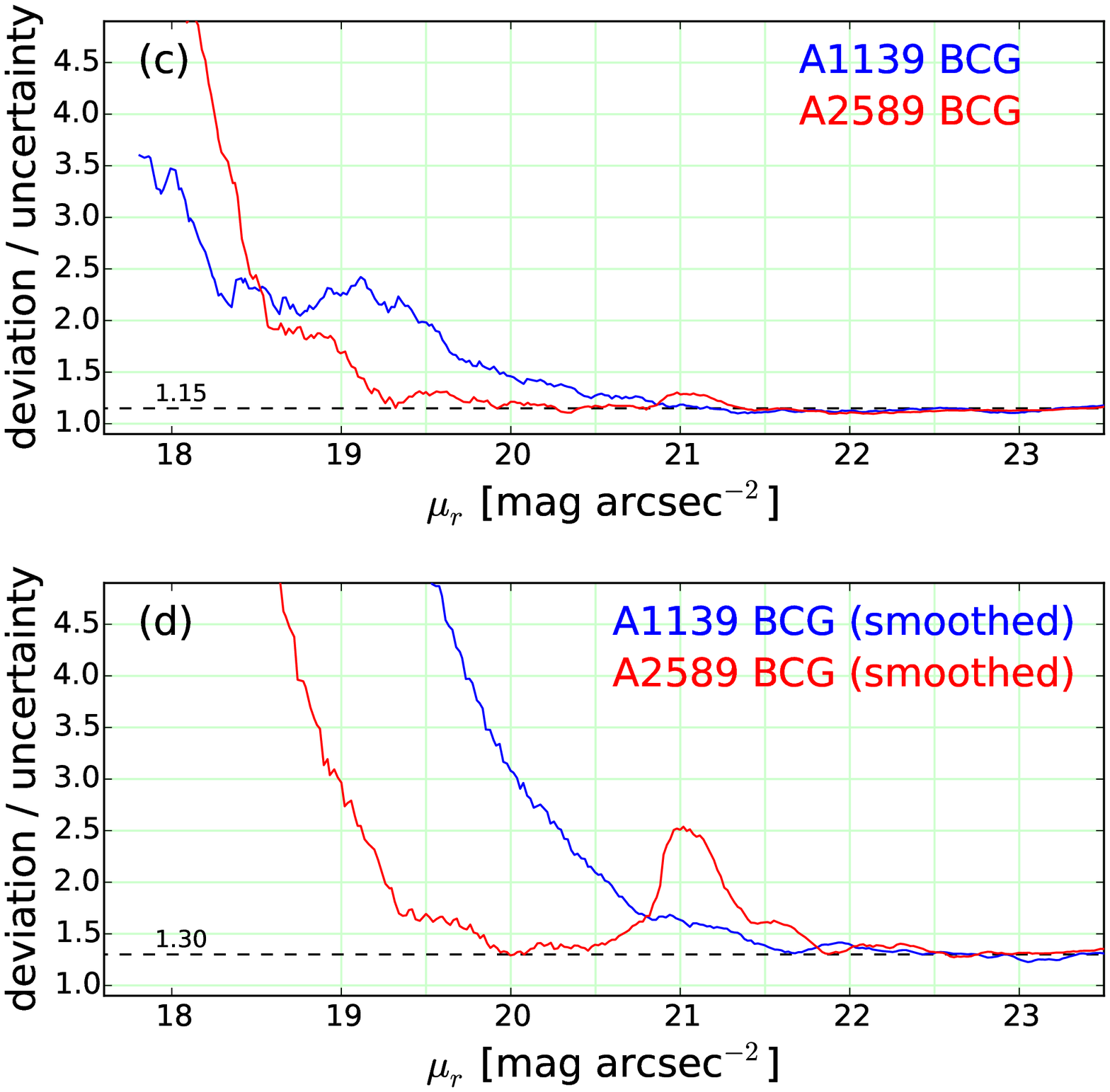}
\caption{(a) and (b): The standard deviations (solid lines) and the single-pixel photometric errors (dashed lines) along $\mu_r$ in the pCMDs. (c) and (d): The ratio between color deviation and photometric uncertainty along $\mu_r$. The ratio converges into a constant value for both of the BCGs as $\mu_r$ increases (horizontally dashed lines; $\sim1.15$ before the smoothing and $\sim1.30$ after the smoothing). The lower panels are the results for the smoothed pCMDs.\label{rips}}
\end{figure*}

In Figure~\ref{slope}, the slope of the {\bcga} backbone becomes smaller (steeper) than $-0.008$ at $\mu_r\sim19.1$ {\umu}, at which the backbone starts to show a recognizable slope\footnote{The slope criterion of 0.008 was empirically chosen, but it approximately corresponds to 3 times of the mean slope-uncertainty ($\approx0.0025$) at $\mu_r~18.8 - 19.3$ {\umu}.}, whereas the slope of the {\bcgb} backbone does at $\mu_r\sim19.7$ {\umu}. 
The difference in the backbone breaks between the BCGs is $\Delta\mu_r\approx0.6$ {\umu}, in the context that the plateau of the {\bcgb} ends at fainter $\mu_r$.
After the pCMD plateaus with different ending points, the backbones bend down (i.e., bluer colors for fainter pixels).
Whereas the $\mu_r$ at which the slope begins is different between the two BCGs, the descending slopes seem to be similar to each other. The details of the slope after plateau for each BCG are summarized in Table~\ref{slopeinfo}. It is found that the two BCGs show similar color variations for $\Delta\mu_r=3.0$ {\umu} after their plateaus, despite the different spatial scales corresponding to the $\Delta\mu_r$.
\citet{mac92} reported that the $g-r$ color profile of the {\bcgb} looks somewhat flat before the outer envelope starts \citep[$\mu_r\lesssim23.8$ {\umu}, based on the Gunn filter system;][]{thu76}, which is not consistent with Figures~\ref{backbone} and \ref{slope}.
Such disagreement may be due to the poor observational conditions in \citet{mac92}, because he used a small telescope (the KPNO Case Western Reserve University Burrel 0.6/0.9-m Schmidt telescope) with low S/N and poor resolution ($2.03''$ per pixel).

\begin{figure*}[!t]
\centering
\includegraphics[width=0.45\textwidth]{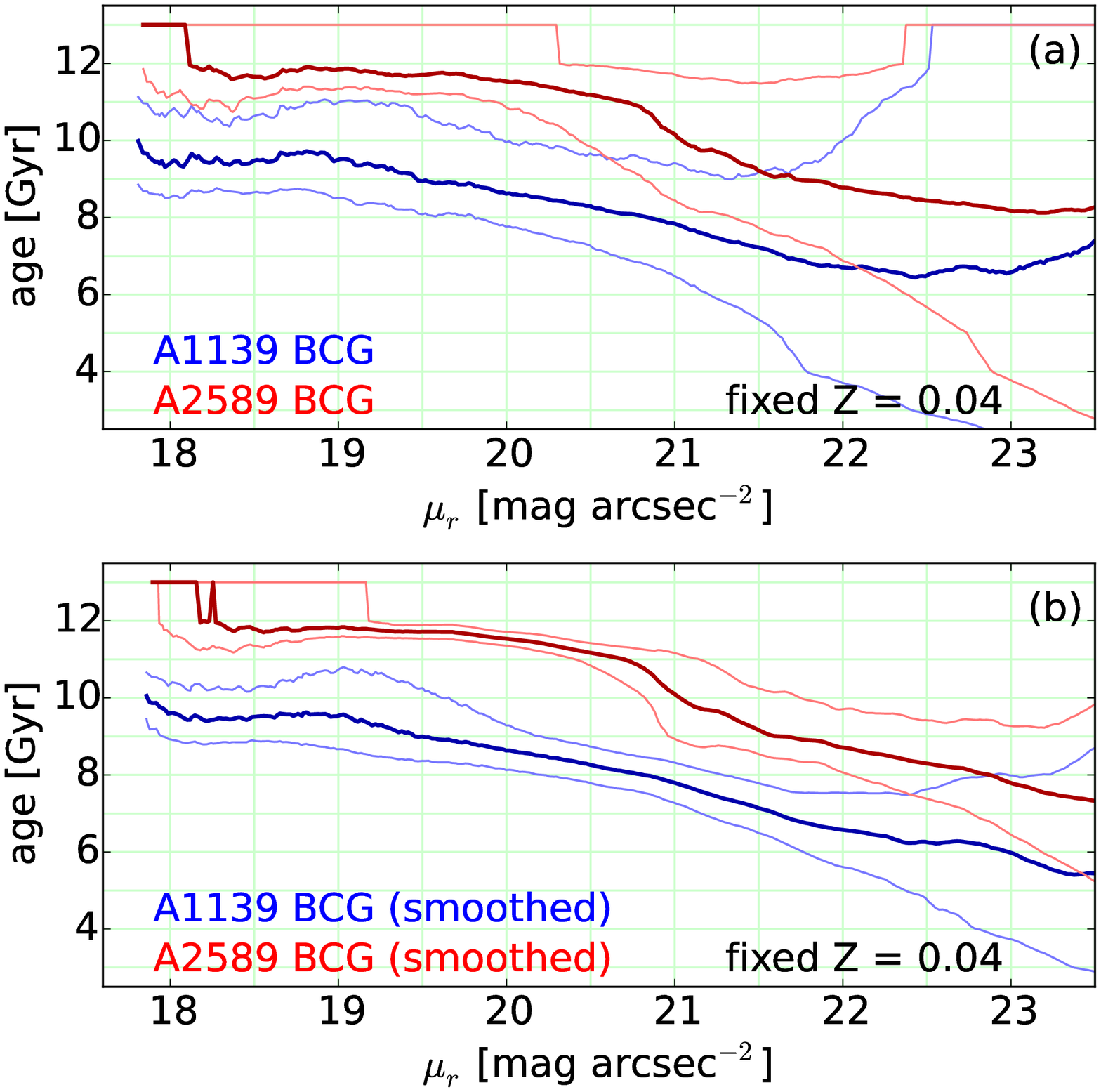}
\includegraphics[width=0.45\textwidth]{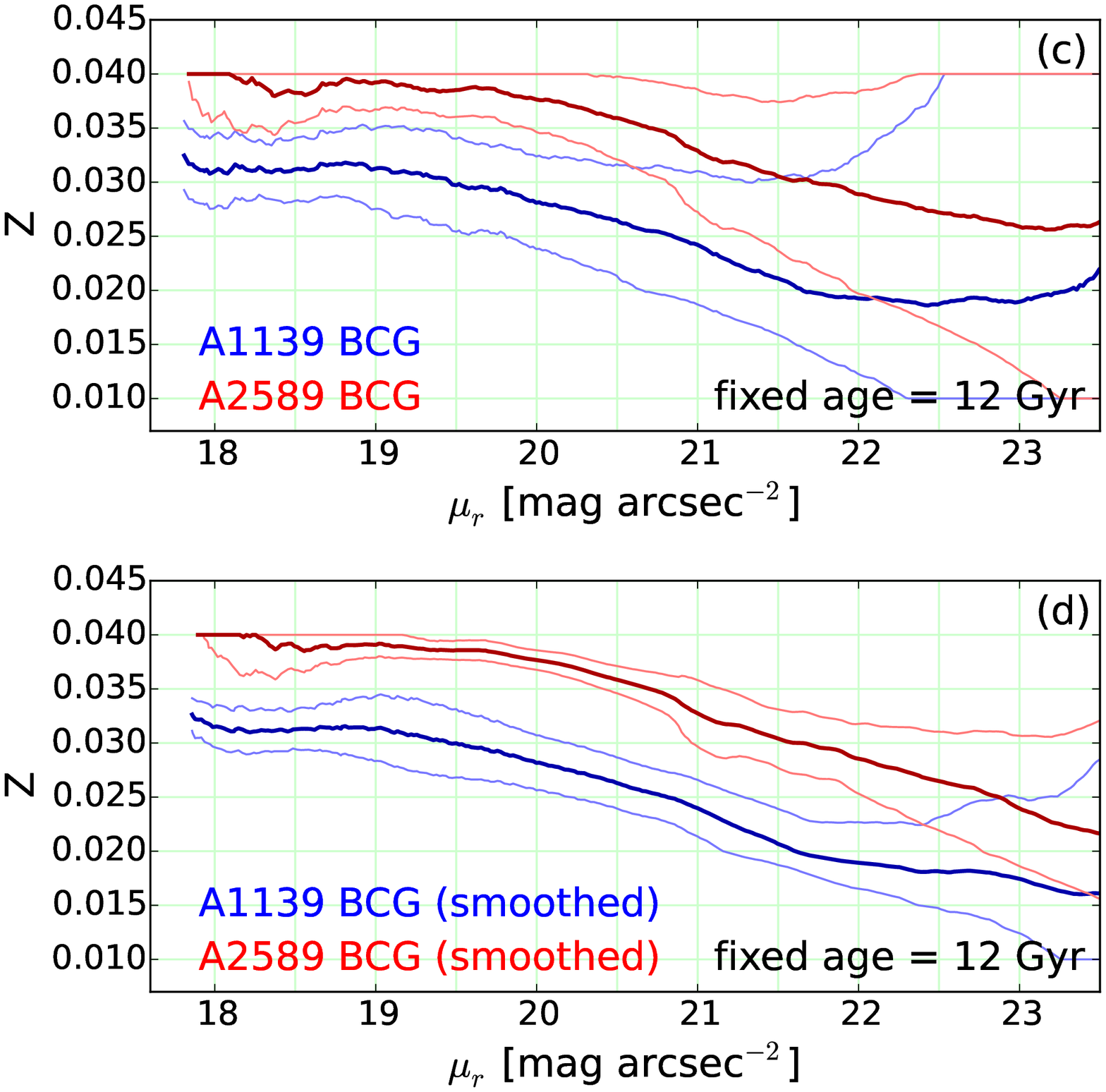}
\caption{(a) and (b): Age conversion of the pCMD backbones at fixed metallicity (Z = 0.04), using composite stellar population models in consideration of gas inflow and outflow. (c) and (d): Metallicity conversion of the pCMD backbones at fixed age (= 12 Gyr), using the same models. The lower panels are for the smoothed pCMDs.\label{agemetal}}
\end{figure*}

\begin{figure*}[!t]
\centering
\includegraphics[width=0.45\textwidth]{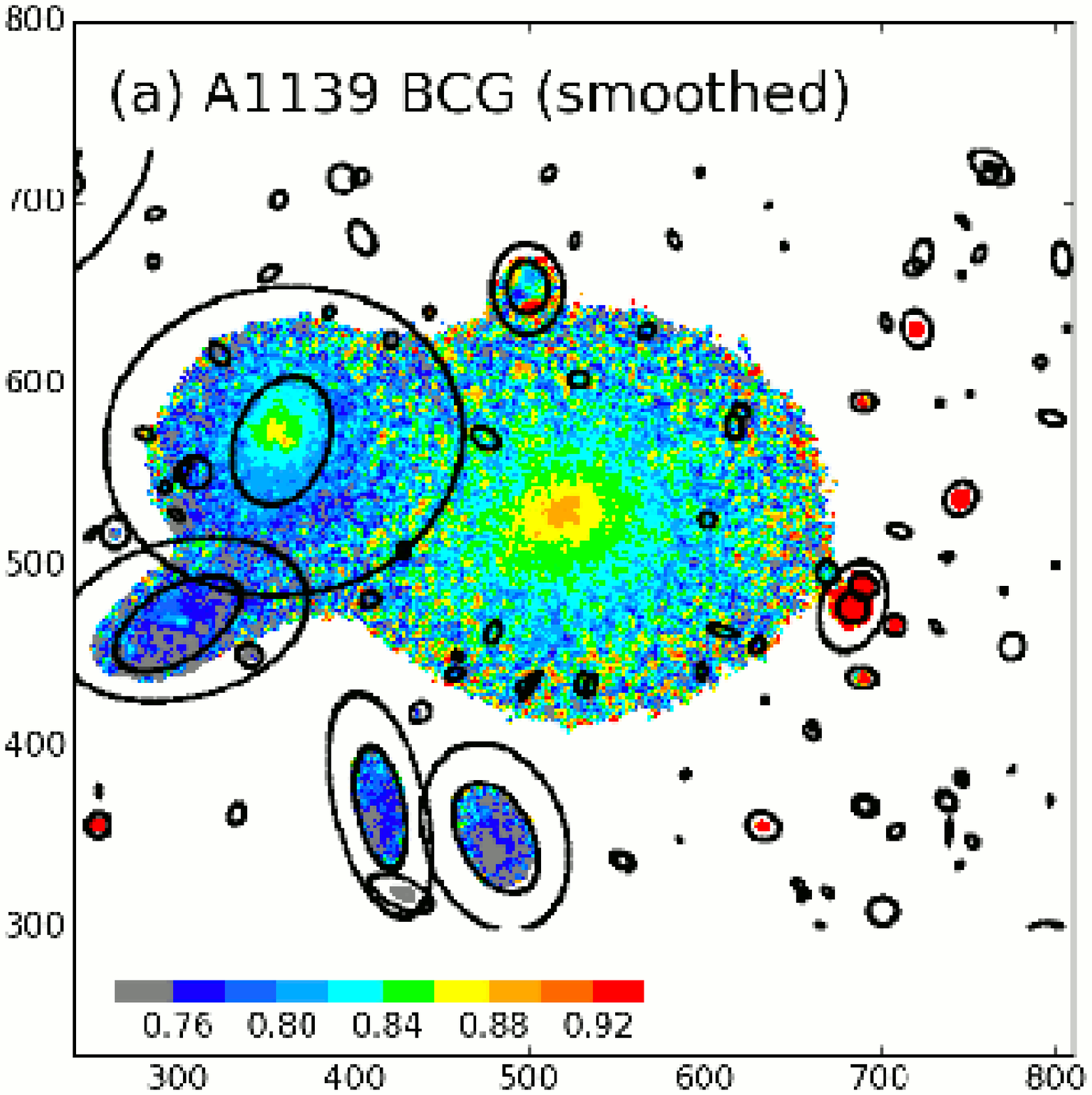}
\includegraphics[width=0.45\textwidth]{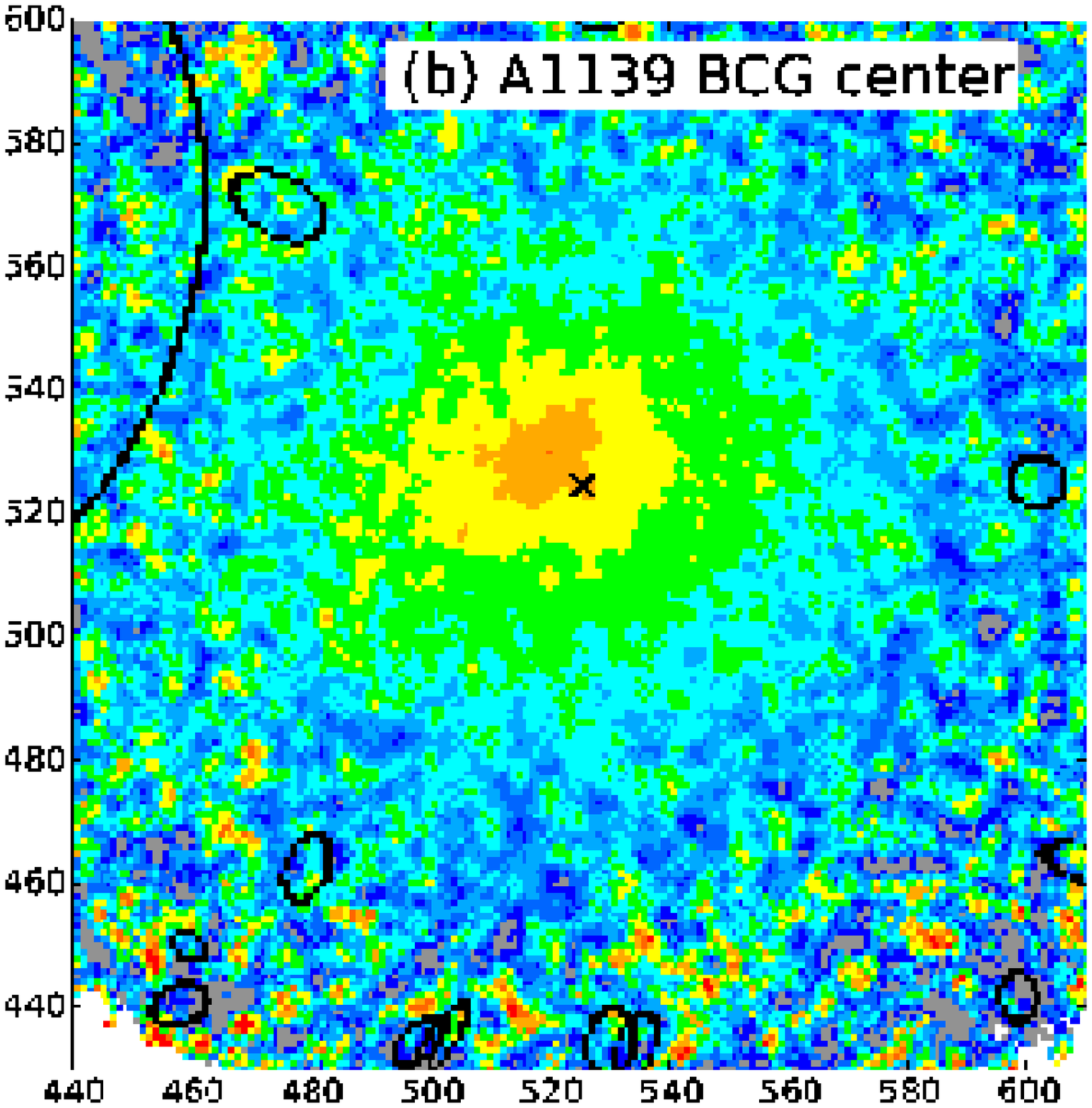}
\includegraphics[width=0.45\textwidth]{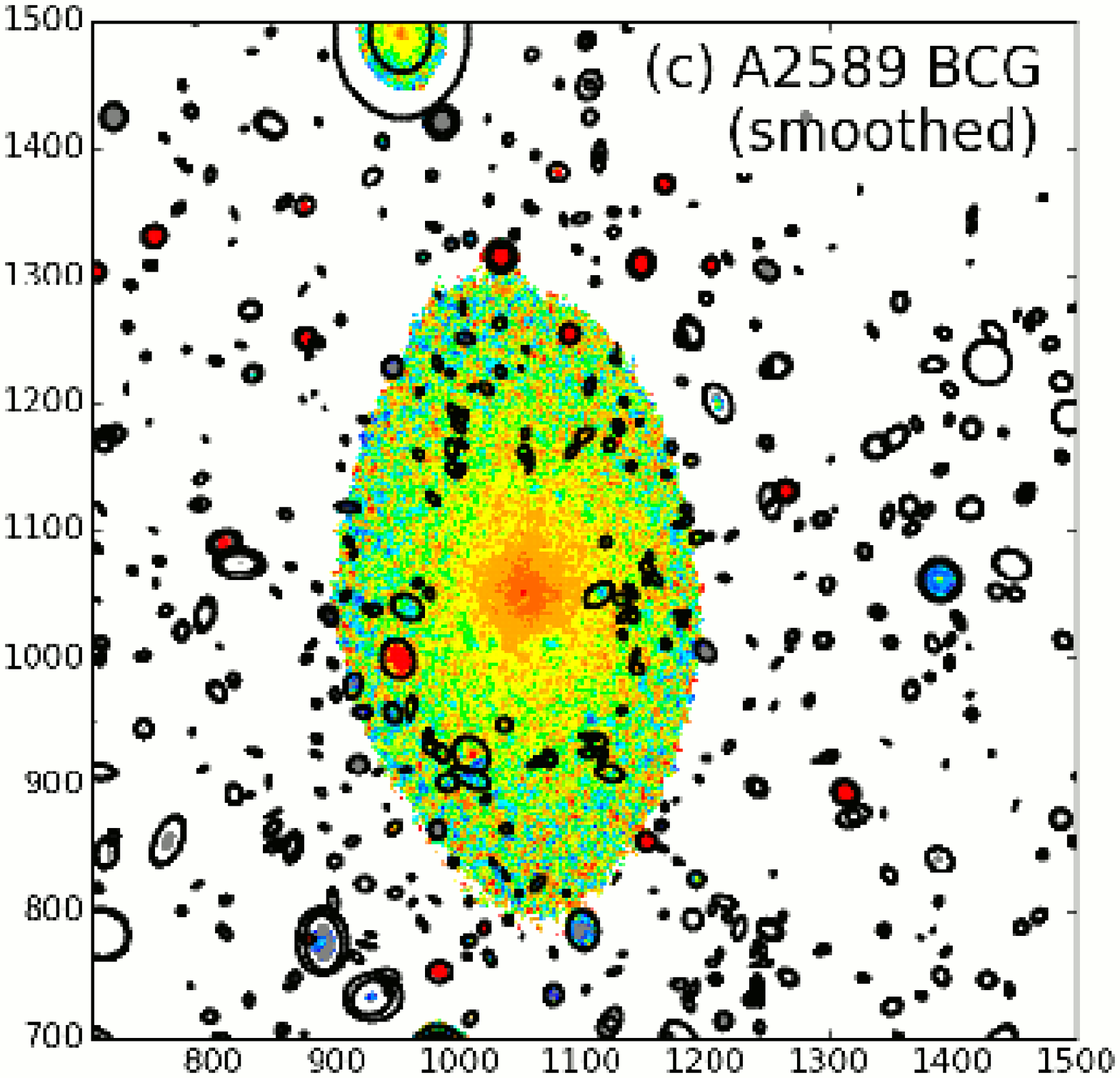}
\includegraphics[width=0.45\textwidth]{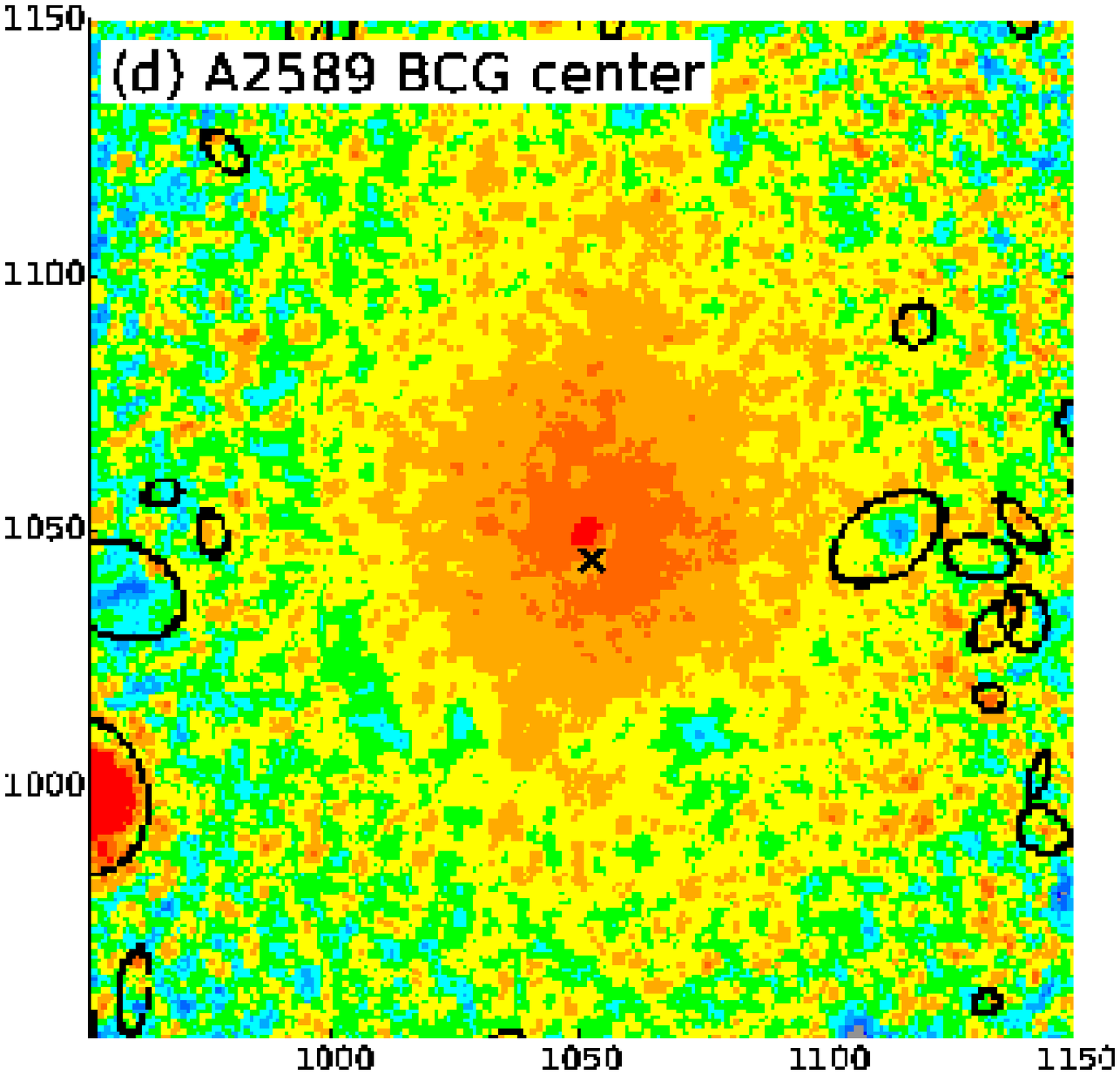}
\caption{The $g-r$ color maps (smoothed; $\mu_r \le 23.5$ {\umu}) of (a) the {\bcga} and (b) its center, and (c) the {\bcgb} and (d) its center. The masked areas are denoted by black ellipses. The BCG center (the mean coordinate of ten brightest pixels) is marked by a black cross in (b) and (d), respectively. The axis labels show the coordinates in pixel, where 100 pixels approximately correspond to the $19''$ or 15 kpc length. \label{colmap}}
\end{figure*}

As revealed more clearly in Figure~\ref{rips}, the pCMD of the {\bcga} has larger color deviations\footnote{Throughout this paper, `color deviation' means the root-mean-square (rms) deviation of color at given $\mu_r$.} at most $\mu_r$ bins, whereas the pCMD color deviation of the {\bcgb} is larger at the brightest end and at around $\mu_r\sim20.9 - 21.2$ {\umu}. 
Note that the comparison of color deviation is only available at $\mu_r\lesssim21.5$ {\umu} (after the smoothing), because the photometric uncertainty becomes more dominant than intrinsic deviation of color for faint pixels. In Figure~\ref{rips}(c) and (d), we present the variation of the ratio between color deviation and photometric uncertainty along $\mu_r$, which converges into the same value for both of the BCGs. The ratio appears to have lower limits of $\sim1.15$ before the smoothing and $\sim1.30$ after the smoothing. If the ratio is close to the limit, the measured color deviation may be dominated by photometric uncertainty, and thus it should be regarded as the upper limit of its intrinsic value. For that reason, the comparison of color deviation between the two BCGs is meaningful only at $\mu_r\lesssim21.5$ {\umu}.
However, even when we limit the comparison to $\mu_r\le21.5$ {\umu}, it is still true that the pCMD of the {\bcgb} is tighter than that of the {\bcga} on average.
 
It is well known that the apparent colors of stellar populations are significantly affected by their ages and metallicities as well as dust reddening. In the optical-band color, however, those three effects are too strongly degenerate to be disentangled only using $g-r$ color information. In Figure~\ref{agemetal}, we simply convert the pCMD backbones into ages or metallicities as a function of $\mu_r$, by fixing one to estimate the other (age = 12 Gyr or Z = 0.04) and by ignoring dust effects. The conversion is based on the composite stellar population models  using the Yonsei-Yale (Y$^2$) isochrones \citep{yi01,kim02,spa13} in consideration of gas inflow and outflow. Since the actual ages and metallicities are expected to vary simultaneously along $\mu_r$, the variations of ages and metallicities after the plateaus summarized in Table~\ref{slopeinfo} may be the upper limits.

\begin{figure}[!t]
\centering
\plotone{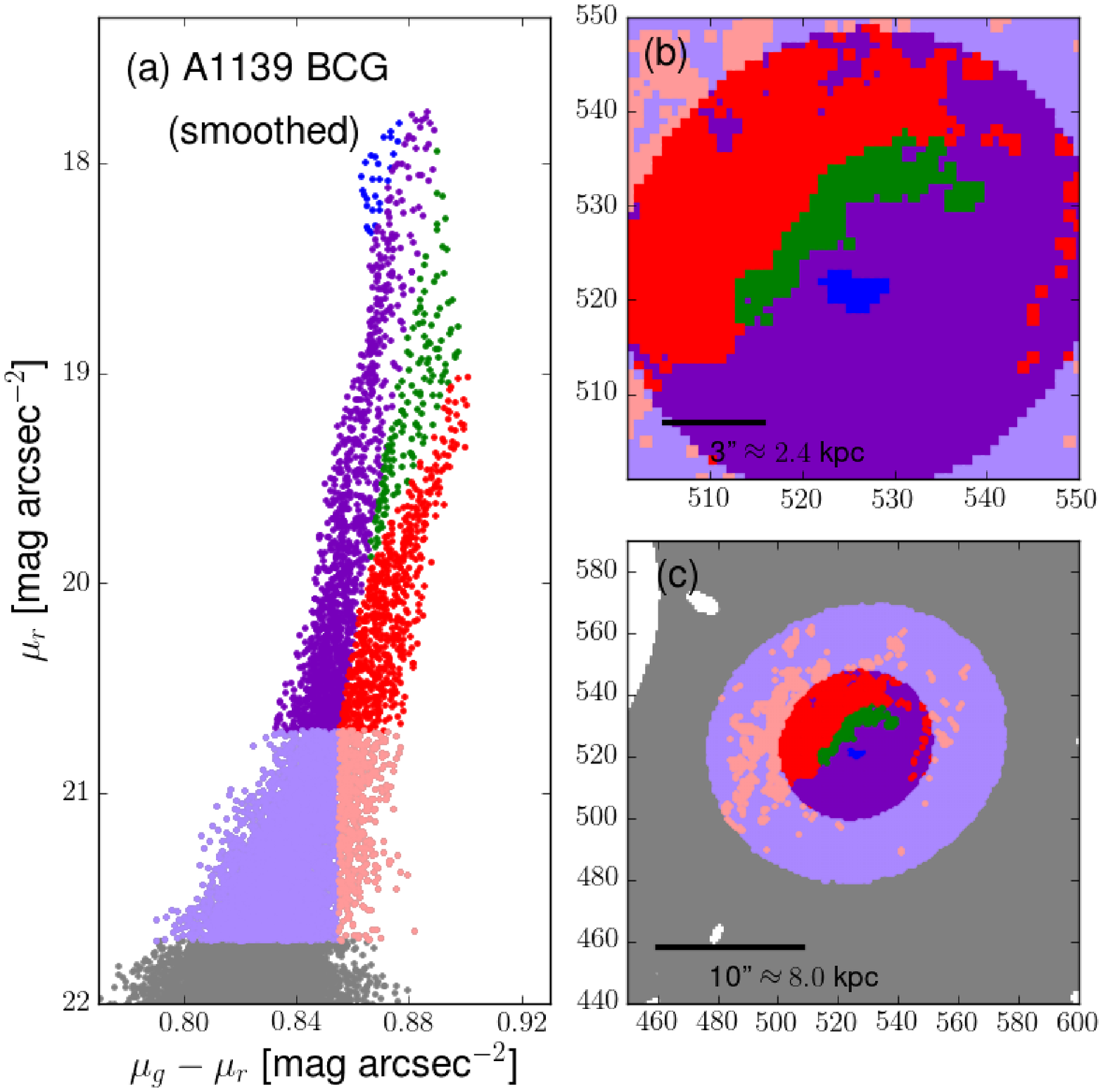}
\caption{(a) The intermediate and bright part of the pCMD for {\bcga}. The colors of the dots are manually coded to distinguish their domains in the pCMD. (b) The spatial distribution of the pixels grouped in their pCMD domains at the small area of the {\bcga} center. The XY-axes show the pixel coordinates in the trimmed image. (c) The same as (b), but the scale is larger.\label{dev1}}
\end{figure}

\begin{figure}[!t]
\centering
\plotone{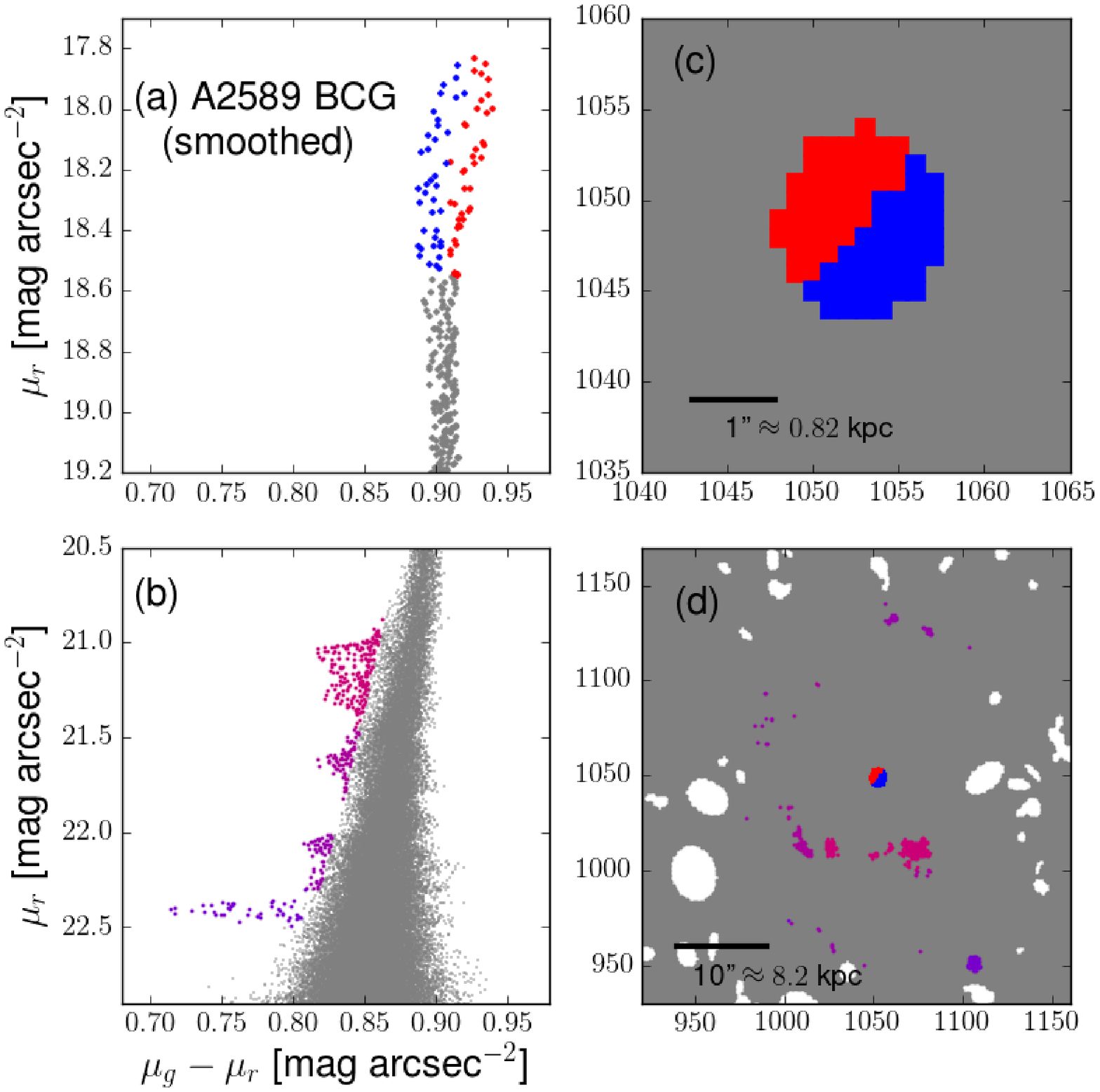}
\caption{(a) The bright part of the pCMD for {\bcgb}. The colors of the dots are manually coded to distinguish their domains in the pCMD. (b) The intermediate and faint part of the pCMD for {\bcgb}. (c) The spatial distribution of the pixels grouped in their pCMD domains at a small area around the {\bcgb} center. The XY-axes show the pixel coordinates in the trimmed image. (d) The same as (c), but the scale is larger.\label{dev2}}
\end{figure}

\begin{figure}[!t]
\centering
\plotone{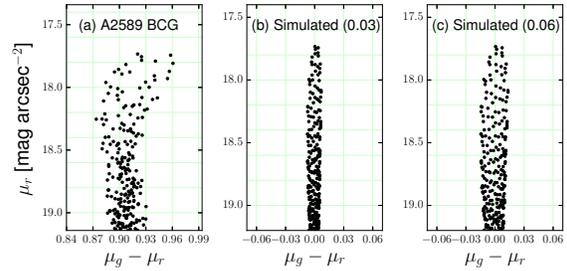}
\caption{(a) The bright part of the pCMD for the {\bcgb} (not smoothed). (b) A simulation of pCMD  produced from the $r$-band image of the {\bcgb} and its shifted image by 0.03 pixel. (c) The same as (b) but the image shift is 0.06 pixel.\label{artshift}}
\end{figure}

\subsection{Fine Features of the pCMDs}\label{finefeat}

\begin{figure}[!t]
\centering
\plotone{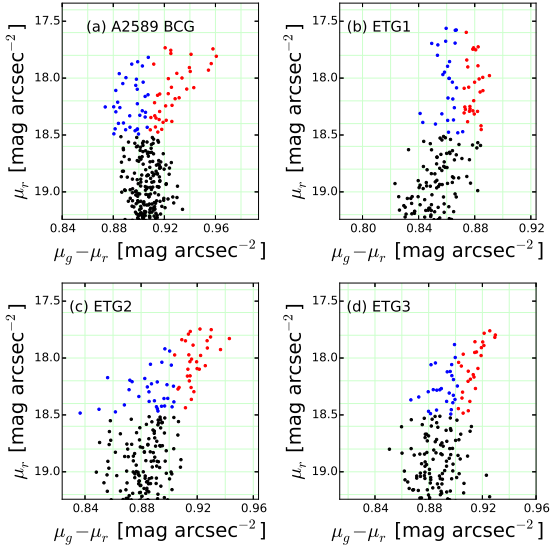}
\caption{ The bright parts of the pCMDs for (a) {\bcgb} and (b) - (d) three bright early-type non-BCGs in A2589. The pCMDs are not smoothed, unlike that in Figure~\ref{dev2}. In each pCMD, the pixels brighter than $\mu_r=18.5$ {\umu} are divided into blue and red ones, the numbers of which are the same as each other.\label{tips1}}
\end{figure}

\begin{figure}[!t]
\centering
\plotone{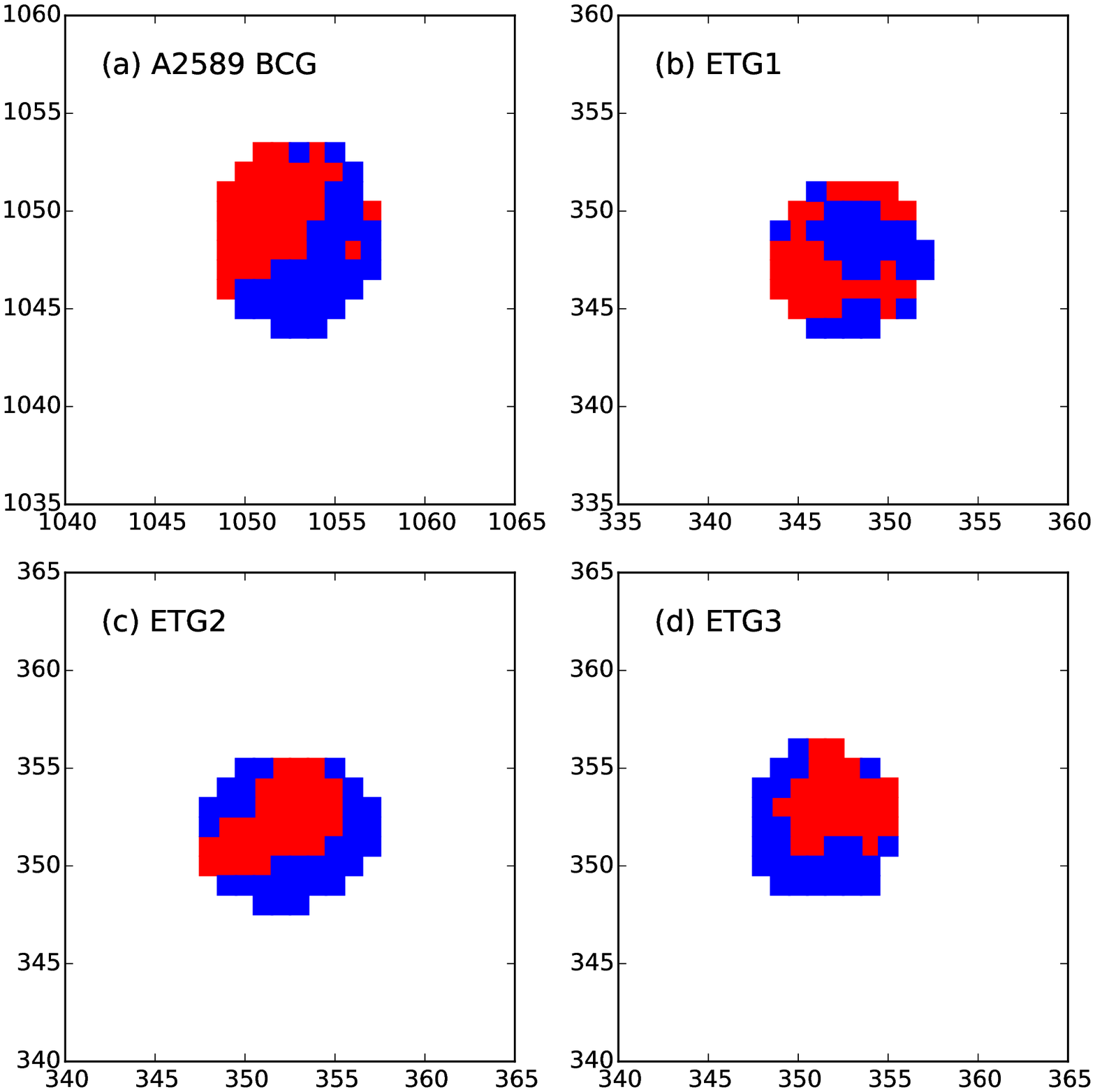}
\caption{The spatial distributions of bright pixels for (a) the {\bcgb} and (b) - (d) three bright early-type non-BCGs in A2589. Blue and red symbols follow the division in Figure~\ref{tips1}.\label{tips2}}
\end{figure}

As mentioned in Section~\ref{anal}, both of the BCGs show noticeable fine features in their pCMDs, which become more obvious after the pixel smoothing. Here, we individually inspect those features, focusing on their spatial distributions.
Before that, we first built $g-r$ color maps to inspect the overall stellar population distributions as displayed in Figure~\ref{colmap}. Both of the BCGs show negative color gradients (redder center and bluer outskirt) as already shown in Figure~\ref{backbone}. Interestingly, in the central regions, both of the BCGs show discrepancies between the peak $\mu_r$ and peak $g-r$ coordinates, which implies that their internal structures may not be quite symmetric. In addition to the core displacement, the color distributions around the cores seems to have some sub-structures, even when we ignore the masked regions.

To investigate such asymmetry and fine structures more systematically, we inspect the spatial distributions of the pixels outlying in the pCMDs.
In Figure~\ref{dev1}, the pCMD fine features of the {\bcga} are examined. In the smoothed version, the {\bcga} shows relatively large dispersion at intermediate and bright parts of its pCMD, which seems to be approximately divided into 3 branches although they are not very distinct from one another. We manually distinguish the 3 branches and one additional  bright blue bump as shown in Figure~\ref{dev1}(a).
Figure~\ref{dev1}(b) and (c) show the spatial distributions of those manually defined fine features with different scales. It appears that the pixels in the red branch are clustered in the upper-left side from the BCG center, which is consistent with the direction to a very close and bright neighbor galaxy (the rank-11 galaxy in A1139; separated from the BCG by $\sim32''$ or 25 kpc).
The masking aperture for the neighbor is partially shown in Figure~\ref{dev1}(c) as the large blank area at the upper-left corner.
The pixels in the intermediate branch (green dots) occupy the intermediate space between the red-branch pixels and the center of the BCG.

The smoothed pCMD of the {\bcgb} shows a tighter sequence than the {\bcga}, except for the brightest-end pixels and a small number of intermediate and faint pixels, as shown in Figure~\ref{dev2}. Since the brightest-end pixels form an elongated ring-like feature in Figure~\ref{dev2}(a), we divide them into two branches.  In addition, the blue-ward outliers in Figure~\ref{dev2}(b) are grouped into 4 subsets. Their spatial distributions are presented in Figure~\ref{dev2}(c) and (d).
The brightest-end pixels form the BCG core, which is spatially divided into two distinct regions. 
However, this kind of a compact and asymmetric feature can be artificially produced if the image alignment is incorrect. Thus, we repetitively checked the image alignment using point sources, and as a result we confirmed that the images in the two bands are correctly aligned to each other within 0.03 pixel ($\sim0.006''$) uncertainty.

To estimate the effect of the alignment uncertainty, we simulate how large color spreads are produced by artificial image shifts.
In Figure~\ref{artshift}, the not-smoothed pCMD of the {\bcgb} is compared with the simulated pCMDs, which are artificially produced from the $r$-band image of the {\bcgb} and its shifted images by 0.03 pixel and by 0.06 pixel respectively. These simulations help us to understand how significantly the alignment uncertainty affects the color spread in a pCMD. The pCMD using 0.03-pixel shift (similar to our image alignment uncertainty) produces small color spread about $\Delta(g-r)\sim0.015$ at $\mu_r=17.9 - 18.0$ {\umu}, which is only one third of that in the original pCMD ($\Delta(g-r)\sim0.075$).
Even in the pCMD using 0.06-pixel shift (twice larger than our image alignment uncertainty), the color spread at $\mu_r=18.0$ {\umu} is $\Delta(g-r)\sim0.050$, which is still considerably smaller than that in the original pCMD.
That is, the alignment uncertainty seems not large enough to explain that the feature in Figure~\ref{dev2}(c) is artificially produced. Thus, it is suspected that there may be genuinely two distinct stellar populations spatially separated at the compact ($\sim1.8''$ or 1.4 kpc in diameter) core region of the {\bcgb}. Otherwise, a very compact dust clump that partially covers of the BCG core is also a possible candidate for the origin of this feature.

\begin{figure*}[!t]
\centering
\includegraphics[width=0.95\textwidth]{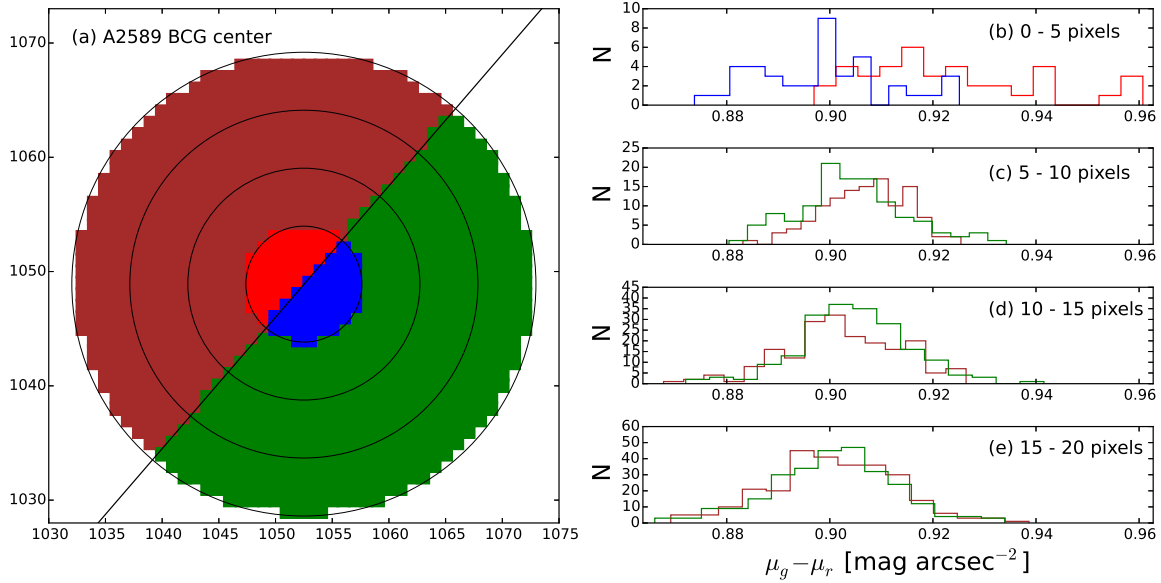}
\caption{(a) The central area division of the {\bcgb}; and the pixel color distributions at given area: (b) between $0-5$ pixels radii, (c) between $5-10$ pixels radii, (d) between $10-15$ pixels radii, and (e) between $15-20$ pixels radii. \label{corecol}}
\end{figure*}

\begin{figure*}[!t]
\centering
\includegraphics[width=0.45\textwidth]{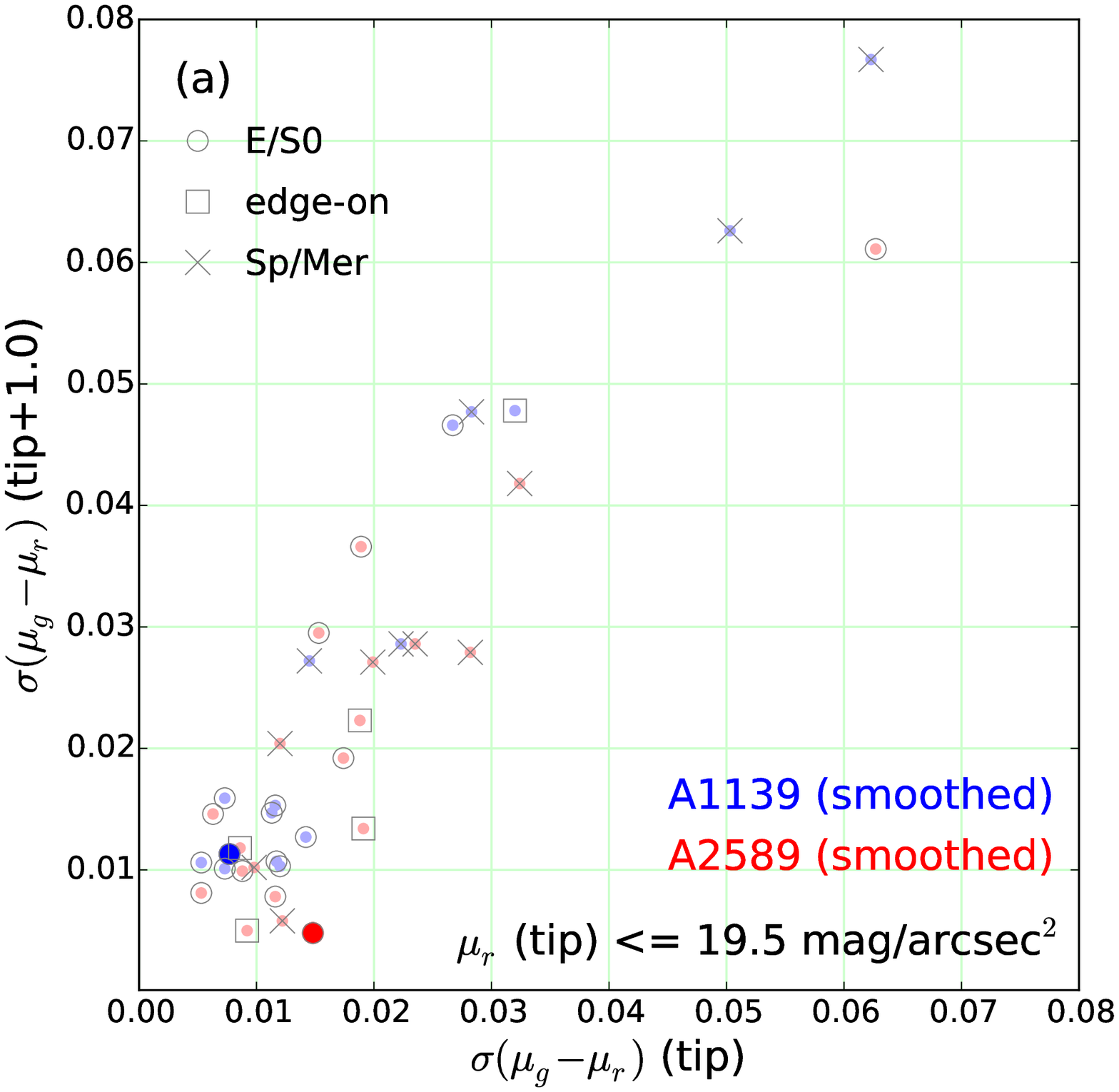}
\includegraphics[width=0.45\textwidth]{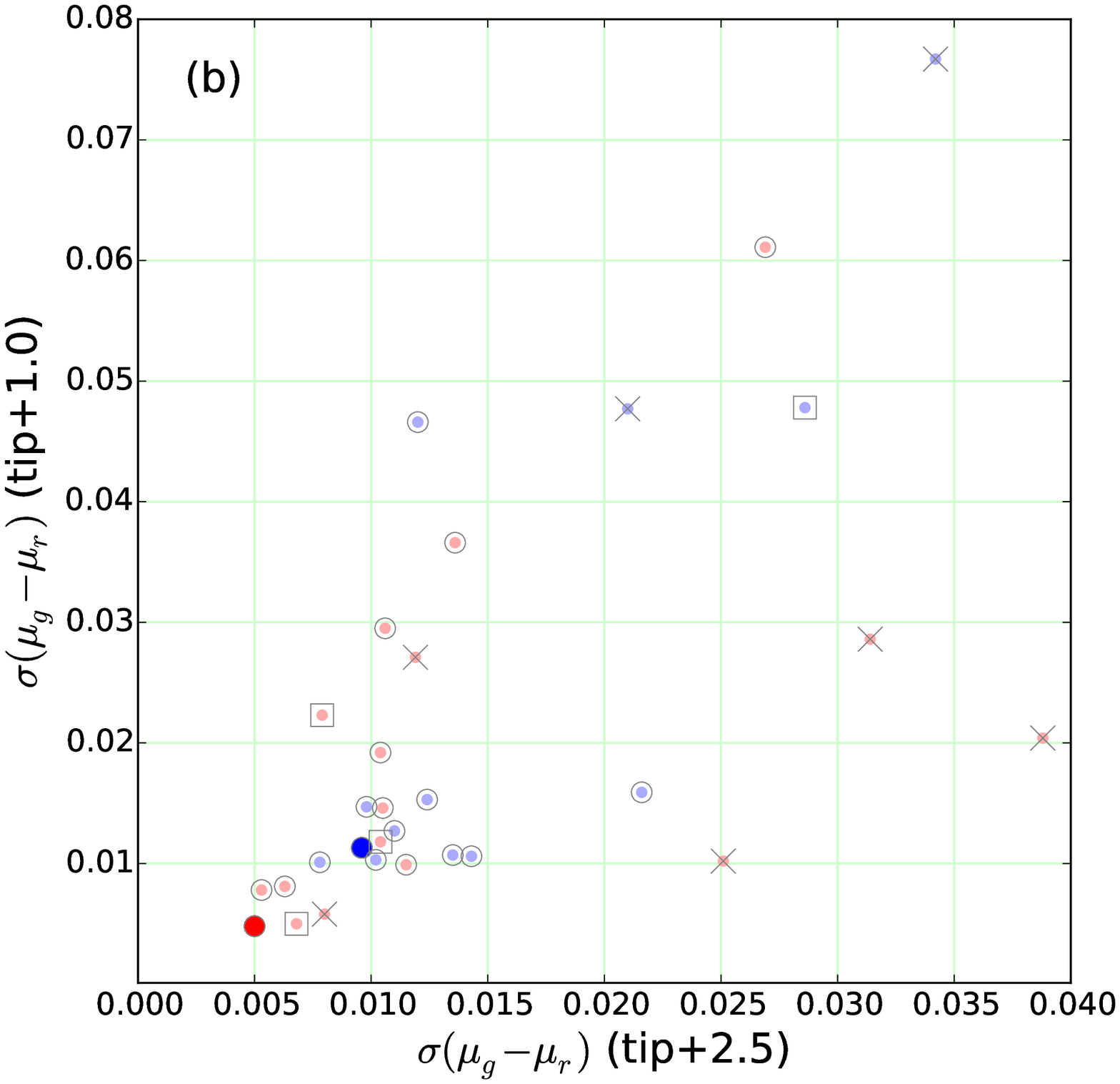}
\caption{Comparison of pCMD color deviations at three $\mu_r$ points ($\mu_r$(tip), $\mu_r$(tip) + 1.0 {\umu}, and $\mu_r$(tip) + 2.5 {\umu}) between BCGs (large and dark filled circles) and non-BCGs (small and light filled circles). The non-BCGs are classified into early-type (open circles), edge-on disk (squares), or spiral/merger (crosses) by eyes. The galaxies with $\mu_r$(tip) $> 19.5$ {\umu} are not plotted because their pCMDs are often significantly contaminated by neighbor objects and thus masked too much. All parameters are based on the smoothed pCMDs.\label{coldev}}
\end{figure*}

To check the reliability of this feature further, Figure~\ref{tips1} compares the bright parts of the not-smoothed pCMDs for the {\bcgb} and the three brightest early-type non-BCGs without morphologically unusual features in A2589. The brightest parts of the pCMDs for the three non-BCGs are found to be less deviated than that for the {\bcgb}. Where we define $\mu_r$(tip) as the surface brightness at the brightest tip of a pCMD, the color spread at [$\mu_r$(tip) : $\mu_r$(tip) + 0.5 {\umu}] is as large as $\Delta(g-r)\sim0.07$ for the {\bcgb}, whereas those for the three non-BCGs are at most $\Delta(g-r)\sim0.03-0.05$.
We also compare the spatial distributions of blue and red pixels between the four galaxies. Here, we do not manually divide the bright pixels into blue and red ones whereas we did in Figure~\ref{dev2}, because the detailed shapes of the brightest end of the pCMDs are so various that manual divisions may be unfair.
Instead, we simply divide the pixels brighter than $\mu_r=18.5$ {\umu} in each pCMD into blue and red ones only based on their $g-r$ color indices, where the $g-r$ criteria are selected for the numbers of blue and red pixels to be the same as each other, as shown in Figure~\ref{tips1}.
Their spatial distributions are compared in Figure~\ref{tips2}. Since all the galaxies have similar alignment uncertainties between their $g$- and $r$-band images, the spatial separation of blue and red pixels would be usually found if it were due to image misalignment. However, as shown in Figure~\ref{tips2}, the spatial separation of blue and red pixels in the three non-BCGs are not as obvious as that in the {\bcgb}, although the ETG3 shows a moderate signature of the blue-red pixel separation.

Finally, we examined the pixel color distribution around the {\bcgb} center. If the core blue-red pixel separation were due to image misalignment, color biases would be detected even out of the core region, although the signal would not be as clear as that at the core. In Figure~\ref{corecol}, the central area of the {\bcgb} is divided into 4 sections: the radius ranges of $0-5$ pixels, $5-10$ pixels, $10-15$ pixels and $15-20$ pixels. The color distributions at the upper-left region and at the lower-right region in each section are compared in Figure~\ref{corecol}(b)-(e), in which color biases are detected within 10 pixel radius. However, such a trend disappears at the radii larger than 10 pixels: the trend even seems to be slightly inverted (i.e., the pixels in the upper-left regions look bluer) in Figure~\ref{corecol}(d) and (e).

Despite these test results, it can not be asserted yet that the spatial separation between blue and red pixels at the center of the {\bcgb} is genuine, because it is hard to predict the small-scale variation of point spread function in the images from ground-based observations. Thus, we cautiously conclude that this may be a real feature, but there still remains a possibility that it is spurious.
The possible origin of such a feature on the assumption that it is real will be shortly discussed in Section~\ref{dis2}.

In addition to the asymmetric core, more unusual features are found at intermediate and faint parts in the pCMD of the {\bcgb}, as shown in Figure~\ref{dev2}(b).
Those blue-ward outliers appear to be gathered along a few stream-like features in Figure~\ref{dev2}(d). Because the pixel areas of those features are very small, they are thought to be the remnants of recent infall of a few low-mass star-forming galaxies rather than resulted from significant mergers.

From the results in Figures~\ref{dev1} and \ref{dev2}, it is inferred that the large color deviation in the pCMD is due to multiple stellar populations  or dust clumps possibly originating from merging events. We speculate that relatively recent mergers tend to increase the overall pixel color deviation, whereas the large color deviation only at the brightest end ({\bcgb}) may have resulted from major dry merger at an early epoch.
If it is true, the color deviation as a function of pixel surface brightness will give hints on the mass assembly histories of galaxies.
In Figure~\ref{coldev}, we compare the color deviations at three selected $\mu_r$ points: $\mu_r$(tip), $\mu_r$(tip) + 1.0 {\umu}, and $\mu_r$(tip) + 2.5 {\umu} for each galaxy. A smaller color deviation indicates a tighter pCMD at given $\mu_r$. 
In this comparison, the {\bcgb} shows the smallest deviations at $\mu_r$(tip) + 1.0 {\umu} and $\mu_r$(tip) + 2.5 {\umu} (Figure~\ref{coldev}(b)), which implies that the {\bcgb} has experienced recent merging or interaction events most rarely, whereas the {\bcga} is similar to typical early-type non-BCGs.\footnote{On the other hand, there are 4 early-type galaxies that show very large color deviations ($\sigma(\mu_g-\mu_r)$(tip+1.0) $\gtrsim 0.03$). These galaxies seem to be surrounded by close companions or to have faint disk components, although clear tidal features or spiral arms do not appear in their $r$-band images. The pCMD properties of non-BCGs will be more deeply investigated in our follow-up studies.}
However, the {\bcgb} has unusually large color deviation at $\mu_r$(tip) (Figure~\ref{coldev}(a)) despite its small color deviation at fainter $\mu_r$. This is a unique feature of the {\bcgb}, and no other bright early-type galaxies in A1139 and A2589 are like it.

\subsection{Pixel Luminosity Functions}\label{plfsec}

\begin{figure}[t]
\centering
\plotone{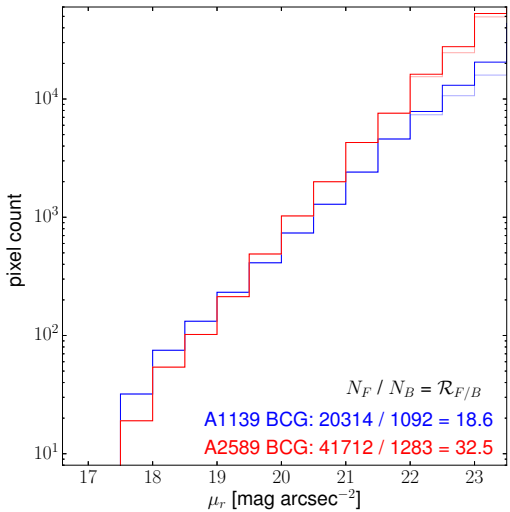}
\caption{Pixel luminosity functions of the BCGs: blue histograms for the {\bcga} and red histograms for the {\bcgb}. The light lines show the histograms before the pixel compensation, while the dark lines present those after the pixel compensation. The faint-to-bright pixel number ratios are denoted (see the main text for details).\label{plf}}
\end{figure}

\begin{figure}[t]
\centering
\plotone{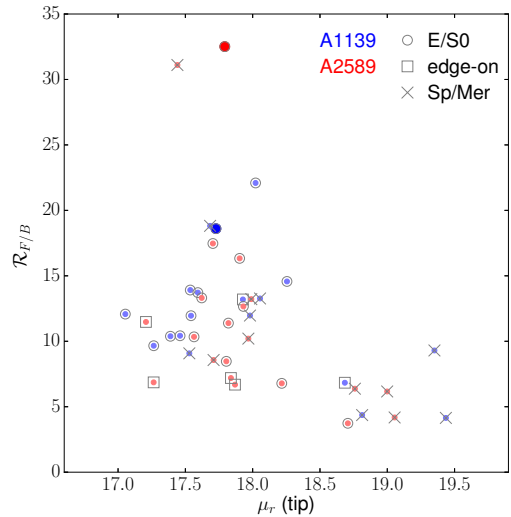}
\caption{Comparison of the faint-to-bright pixel number ratios between BCGs (large and dark filled circles) and non-BCGs.\label{bfr}}
\end{figure}

Finally, we compare the pixel luminosity functions (pLFs) of the two BCGs.
Because many companion and background objects are masked in our pCMD analysis, compensation for the masked pixels is necessary before deriving the pLFs. For a masked pixel, we simply assigned the median $\mu_r$ of the unmasked pixels on the isophotal ellipse, on which the masked pixel lies. 
Figure~\ref{plf} shows the pLFs before and after the compensation of masked pixels. The difference by the compensation is considerable only at $\mu_r>22$ {\umu}.

Despite the good agreement in $\mu_r$(tip) between the two BCGs ($\sim17.7 - 17.8$ {\umu}), the pLF slope of the {\bcgb} is steeper than that of the {\bcga}, as shown in Figure~\ref{plf}.
For quantitative comparison, we define a parameter named faint-to-bright pixel number ratio ($\mathcal{R}_{F/B}$) $\equiv N_F/N_B$, where $N_F$ is the number of pixels with $\mu_r$(tip) + 2.5 {\umu} $\le \mu_r < \mu_r$(tip) + 5.0 {\umu} while $N_B$ is the number of pixels with $\mu_r$(tip) $\le \mu_r \le \mu_r$(tip) $+2.5$ {\umu}.
For the {\bcgb}, $\mathcal{R}_{F/B}$ is as large as 32.5, which is almost twice of the $\mathcal{R}_{F/B}$ of the {\bcga} (= 18.6).

We also compare the $\mathcal{R}_{F/B}$ values of the two BCGs with those of non-BCGs in Figure~\ref{bfr}. It is noted that the BCGs do not have particularly brighter $\mu_r$(tip) compared to non-BCGs, which indicates that the high total luminosities of the BCGs are mainly due to their large outer bodies.
Both of the two BCGs have large $\mathcal{R}_{F/B}$ compared to early-type non-BCGs, probably because BCGs usually have well-developed outer bodies (possibly connected to envelopes) unlike typical early-type galaxies. However, the $\mathcal{R}_{F/B}$ of the {\bcga} is not distinctively larger than early-type non-BCGs and even one early-type non-BCG shows a larger $\mathcal{R}_{F/B}$ than the {\bcga}. On the other hand, the {\bcgb} is distinctive from early-type non-BCGs in Figure~\ref{bfr}. That is, the outer body of the {\bcgb} appears to be very well developed even when compared with the {\bcga}. This results in the similarity in total color between the two BCGs despite the color difference at given $\mu_r$ as mentioned in Section~\ref{backb}, because the outer parts of these BCGs tend to have relatively blue colors.

\section{DISCUSSION}

\subsection{Formation and Growth of the BCGs}

Many recent studies reported that much evidence supports the inside-out formation of massive galaxies \citep[e.g.,][]{van10,pat13,bai14,del16,liu16}. 
Furthermore, a massive galaxy is often thought to form in 2 steps: (1) violent formation of a compact body at an early epoch and (2) later steady size-growth \citep[two-phase formation scenario; e.g.,][]{ose10,hua13,lee13,ras14,pas15,hua16}.
The pCMDs of our target BCGs appear to be consistent with the inside-out two-phase formation, because their backbones show breaks at intermediate $\mu_r$ and the pixels brighter than the breaks (plateau parts; spatially inner regions) tend to be redder (i.e., older or more metal-rich stellar populations) than the pixels fainter than the breaks (sloped parts; spatially outer regions), as shown in Figures~\ref{backbone} and \ref{agemetal}.
However, the details of their formation histories seem to be different from each other, which are reflected in the different curvatures of the pCMD backbones.
Although the pCMD backbones of the BCGs commonly show plateaus at bright parts, their $\mu_r$ ranges are different. For the {\bcga}, the backbone starts to have significantly negative slopes from $\mu_r\sim19.1$ {\umu} (corresponding to $1.7''$ or 1.3 kpc in radius), while it does from  $\mu_r\sim19.7$ {\umu} (corresponding to $2.4''$ or 2.0 kpc in radius) for the {\bcgb}.

From the plateaus found in both of the pCMDs, the central regions of the BCGs are thought to have formed through major dry mergers, which is known to cause a flat color gradient in a galaxy \citep{whi80,ko05,dim09}. 
Combined with the inside-out formation, this result supports a scenario in which the BCGs formed their central bodies through gas-poor mergers at relatively early epochs. The mean stellar age in the central regions are approximated to be as old as 9.4 -- 11.7 Gyr \footnote{We remind that these values were not rigorously determined but simply converted from the $g-r$ color.}. However, this does not mean that the major dry merger events happened at those epochs, because dry mergers hardly trigger star formation activities by definition. A plausible scenario is that strong starburst formed most stars in the currently central bodies at 9.4 -- 11.7 Gyr ago, possibly by gas-rich mergers or large gas inflow, and major dry mergers occurred  after the star formation was quenched (i.e., later than the lookback time of 9.4 -- 11.7 Gyr ).

In this inference, the sizes of the BCGs were different already when they formed in the beginning: the {\bcga} was more compact than the {\bcgb} from the outset. It is noted that the luminosity of the central region corresponding to the pCMD plateau ($\equiv L_{\textrm{\scriptsize plateau}}$) of the {\bcga} ($M_r\sim-20.06$ at $\mu_r\le 19.1$ {\umu}) is lower than $L_{\textrm{\scriptsize plateau}}$ of the {\bcgb} ($M_r\sim-20.35$ at $\mu_r\le 19.7$ {\umu}), but the fractional difference in $L_{\textrm{\scriptsize plateau}}$ is not as large as the fractional difference in their sizes (1.3 kpc versus 2.0 kpc). The ratio of $L_{\textrm{\scriptsize plateau}}$ is only 1.31 ({\bcgb} to {\bcga}), whereas the size ratio is 1.54 and thus the volume ratio is as large as 3.64.
In short, the early body of the {\bcga} may have been fainter and smaller than that of the {\bcgb}, but the stellar number density may have been higher. The different cluster environments may be responsible for such different early properties.

After the early formation of central bodies, the BCGs seem to have grown through minor mergers, forming the sloped parts of their pCMD backbones. In our results, the inside-out formation and galaxy downsizing scenarios make sense, because the galaxies accreting more lately (and thus forming outer regions of the BCGs) may have lower masses and metallicities. Interestingly, despite the different sizes of the plateaus, the faint-side slopes of the backbones are very similar: $-\Delta(g-r)/\Delta\mu_r= 0.017-0.018$.
This can not be simply interpreted as that the growth rates of the BCGs after their early formations are consistent, because the extended radii corresponding to the $\Delta\mu_r = 3.0$ {\umu} are different: 7.3 kpc versus 16.2 kpc, or fractionally 6.4 versus 9.3 times.
On the other hand, when we consider that $\mu_r$ can be a proxy of $R/R_e$, this result indicates that the BCG growth rate may be proportional to the BCG size (or mass).
That is, the homology of the faint and sloped parts of the pCMD backbones supports the scaled similarity in the late formation process of the BCGs.

However, it is still unclear why the fainter pixels should have bluer colors on average if they were formed by minor mergers. Does it indicate that the stellar mass of an infalling galaxy determines the final mean surface brightness (or stellar mass density) of the stars from the galaxy? In other words, do the stars from more massive (and thus older and more metal-rich) galaxies tend to be more densely gathered and thus produce higher mean surface brightness after relaxation? Currently, we do not have a clear answer for this question. Numerical simulations with sufficiently high resolutions will be helpful to address this issue, by tracking back the origin of each mass particle at a given radius.

\subsection{Evidence of Merging and Interaction}\label{dis2}

One of the outstanding differences between the BCGs is the color deviation at given $\mu_r$. It is larger in the {\bcga} at most $\mu_r$ bins, which indicates that the recent growth of the {\bcga} has been more active, possibly due to the later start of growth than the {\bcgb}. Meanwhile, the {\bcgb} also has some evidence of recent minor interactions or mergers, showing several trails possibly from low-mass star-forming galaxies fallen into the BCG. However, the fraction of the pCMD outliers in the {\bcgb} is much smaller than that in the {\bcga}. This implies that the {\bcgb} has not been completely unperturbed, but the recent perturbations were very small.
On the other hand, the {\bcga} shows evidence of interaction with a bright and close neighbor galaxy on its eastern side. The red colors of the pCMD outliers at intermediate brightness may be due to different (older or more metal-rich) stellar populations or due to dust reddening.
Whichever is right, they are considerably biased to the direction of the neighbor galaxy, and thus their origin seems to be closely related with the neighbor galaxy.

In Section~\ref{data}, we mentioned a possibility that several bright galaxies in A1139 may merge into a single massive BCG in the future. Here, we speculate how the pCMD of the {\bcga} will change if it comes true.
The $m_{12}$ value for A1139 is only 0.63, but the bright galaxies from rank-2 to rank-5 in A1139 are separated from the BCG by considerable distances ($\gtrsim10'$ or 500 kpc). Thus, equal-mass ($<$ 2:1 in mass ratio) mergers are hardly expected to happen in near future. However, two bright galaxies are very close to the BCG: the rank-10 and rank-11 galaxies are located within 60 kpc ($\approx76''$) from the BCG, each of which is about $40\%$ of the BCG in luminosity. It is probable that these galaxies will merge into the BCG in the future, since they already show evidence of ongoing interactions with the BCG.
If the mass-to-light ratio is not so different between the BCG and the non-BCGs, the merging with the two close galaxies will be major mergers, which are often defined as mergers with mass ratio smaller than 4:1 \citep[e.g.,][]{cas14}.
Then, the pCMD of the {\bcga} will suffer significant changes because the stellar orbits will be largely mixed and redistributed \citep[e.g.,][]{dim09}. After relaxation, the pCMD will have  a backbone with a larger plateau, the color of which may become slightly bluer than the current one, due to the mixture with bluer stellar populations that are currently at outer regions of the BCG. At that time, compared to the {\bcgb}, the pCMD of the {\bcga} is expected to have bluer mean color, larger color deviation and a larger plateau.

On the assumption that larger color deviations at given $\mu_r$ result from recent tidal interactions or merging events, Figure~\ref{coldev}(b) indicates that any bright galaxies in the two clusters are not as unperturbed as the {\bcgb}. However, despite the tightness of the pCMD at most $\mu_r$ bins, {\bcgb} shows a peculiar feature of double sequences at the brightest tip. 
The pixels in the double sequences form a compact core that is divided into two spatially distinct regions. We checked the reliability of this feature in several ways, but we could not explicitly reject the possibility that it is a real feature. Thus, here we discuss the possible origin of the core asymmetry in the {\bcgb}, on the pure supposition that it is genuine.

Although we coded the pixels with blue and red colors for visibility in Figure~\ref{dev2}, the actual color difference between the two sequences is within $\Delta(g-r)=0.05$ and the stellar populations in there are estimated to be old ($>11$ Gyr) and metal-rich ($Z>0.035$) as shown in Figure~\ref{agemetal}.
As we discussed previously, the central bodies of the BCGs are thought to form through major dry mergers at early epochs, and thus the asymmetric compact core of the {\bcgb} may be the remnant of such a major dry merging event between two old and metal-rich objects long time ago. 
The problem is how such two adjacent stellar systems keep separated for a long time (probably several Gyr) at the center of a massive BCG with very high stellar velocity dispersion. If the core of the BCG is purely dispersion supported, then the stellar systems will be rapidly mixed up, probably in a few Myr.

As a possibility, it can be suspected that the {\bcgb} core may be kinematically decoupled \citep[a kinematically distinct core, KDC;][]{hau94,mcd06}. In many previous studies, it was argued that galaxy-galaxy major mergers can produce some distorted central structures of stellar orbits \citep[e.g.,][]{her91,bal98,hof10} and many early-type galaxies were reported to often host KDCs \citep[e.g.,][]{dav01,ems04,kra11,ems14}.
However, even if the {\bcgb} has a KDC, the core system with separated populations can not be sustained if it follows the Keplerian rotation, in which just a single rotation may perfectly mix up the separated populations. Thus, the only possibility is the case that the core rotates like a bulge's center in a spiral galaxy (e.g., the Milky-Way bulge), which is known to be almost solid-body-like within $\sim1$ kpc diameter from the center.
We emphasize that this is a pure speculation with little evidence, based on the assumption that the asymmetric core of the {\bcgb} is a genuine feature. There is still a considerable possibility that it is just a spurious result. Higher-resolution multi-band imaging and integral field spectroscopy of the {\bcgb} core region will be useful to confirm this issue.

Figure~\ref{coldev}(a) shows the uniqueness of the {\bcgb} in the backbone tightness: only the {\bcgb} has the tight pCMD at intermediate and faint $\mu_r$ but large color deviation at the bright-end at the same time.
The uniqueness of the {\bcgb} also appears in its pLF and $\mathcal{R}_{F/B}$ ratio as shown in Figure~\ref{bfr}.
Compared to the {\bcga}, the {\bcgb} has a twice larger $\mathcal{R}_{F/B}$, and no other early-type non-BCGs are comparable with it.
Since recent studies \citep[e.g.,][]{van10} and our results in Section~\ref{backb} support that the outer body of a bright galaxy has grown via progressive minor mergers, the distinctively large $\mathcal{R}_{F/B}$ of the {\bcgb} indicates that the BCG has experienced more minor mergers than the {\bcga} as well as any other non-BCGs.

\subsection{Caveats}

We discuss several caveats in our results and their interpretations.
First, we masked many companion and background objects around the BCGs, but it is sometimes difficult to distinguish sub-structures of a BCG from its faint companions. Due to the technical limit, there may be missing contaminants particularly in the bright and crowded area at the center of a galaxy cluster. Furthermore, this is not only a technical problem but also an intrinsic problem, because the sub-structures of a BCG are thought to often originate from accreted satellite galaxies. That is, there are no clean divisions among satellite galaxies, tidal debris, and BCG sub-structures. 

Since we used the SExtractor for the detection of the objects to be masked, objects sustaining their original shapes may have been detected probably, but significantly disassembled objects may have been missed.
However, such significantly disassembled objects are intrinsically not distinguishable from sub-structures of a BCG.
In our results, the {\bcga} pCMD has no blue-ward outliers at $\mu_r \sim 21.0 - 22.5$ {\umu}, at which the {\bcgb} pCMD shows several obvious blue plumes. Note that the {\bcga} has very a few small masks at the regions corresponding to the $\mu_r$ range (Figure~\ref{dev1}). That is, the features of blue-ward outliers found in the {\bcgb} pCMD (possibly originating from disassembled accreting satellites) do not exist in the {\bcga}.

Second, the pCMD backbone curvature of the {\bcga} may not perfectly reflect its mass assembly history, because its pCMD has some branches at intermediate surface brightness, probably originating from current interactions with a close neighbor galaxy.
In Figure~\ref{dev1}, such branches are distinguished manually, which seems to make the `intrinsic backbone' have a shorter intrinsic plateau. If we estimate the pCMD backbone and its curvature only using the main branch (blue dots in Figure~\ref{dev1}(a)), the $\mu_r$ at which the plateau ends will change from $\sim19.1$ {\umu} to $\sim18.8$ {\umu}, which corresponds to $\sim1.3''$ or $\sim1.1$ kpc in radius.
In this case, the physical scale of the core region with flat color gradient in the {\bcga} becomes even smaller, indicating that a smaller body has formed by early major dry mergers. However, it does not change our conclusion: in the dynamically younger cluster A1139, a smaller central body of the BCG formed at an early epoch.

Third, the {\bcgb} is brighter and more massive than the {\bcga}, as shown in Table~\ref{clinfo}. Thus, we need to be cautious when connecting the properties of the BCGs with their host clusters, because the mass of a galaxy is an important driver of galaxy evolution as the environment is \citep[e.g.,][]{par07,lee10}. Thus, some differences between the two BCGs may result from their mass difference rather than environmental difference. For example, the central body of the {\bcgb} that probably formed at an early epoch is larger than that of the {\bcga}. About that, we inferred that it may be related with the early formation of A2589, but it may be just because a more massive galaxy tends to form earlier \citep{cow96,tre05}.
Similarly, the difference in stellar populations between the BCGs may be also due to their mass difference rather than environmental difference, because massive galaxies are known to be more metal-rich \citep{tre04,bro07a}.

It is not easy to address this issue, since more evolved clusters (e.g., fossil clusters) tend to have more massive BCGs \citep{men12,wen15}. The best way is to compare BCGs with similar masses hosted by clusters in different evolutionary stages or vice versa, but it is difficult to secure such a perfect sample of BCGs and clusters with good data quality and consistent observational conditions.
Nevertheless, to disentangle the cluster environmental effect from the effect of the BCG mass, a much larger sample needs to be investigated in the future, including galaxy clusters in various evolutionary stages but with BCGs in similar mass ranges.
Here, we simply regard mass as one of the properties of a BCG that are affected by the evolutionary stage of its host cluster. In other words, from our results, the BCG in a dynamically more relaxed cluster tends to have formed a larger central body through major dry mergers at an early epoch and finally evolves to be more massive.

\section{CONCLUSION}

We conducted a case study to understand the coevolution of BCGs and their host clusters.
The BCGs in dynamically young and old clusters A1139 and A2589 were compared using the pCMD analysis method. We found that:
\begin{enumerate}
\item Both of the BCGs show pCMD shapes similar to those of typical early-type galaxies. However, at given surface brightness, the pCMD of the {\bcgb} has redder mean pixel color ($\Delta(g-r)\approx0.02-0.04$) and smaller pixel color deviation than the {\bcga}. This indicates that the BCG in the dynamically more relaxed cluster A2589 has older and/or more metal-rich stellar populations that are spatially better mixed.
\item In the pCMD backbones, the {\bcgb} has a plateau ending at fainter $\mu_r$, which indicates that the {\bcgb} formed a larger central body probably through major dry mergers at an early epoch ($\sim2.0$ kpc in radius) than the {\bcga} with a central body of $\sim1.3$ kpc in radius (if only the main branch is considered, $\sim1.1$ kpc). Their faint outer regions seem to have grown by subsequent minor mergers after the early formation, which is consistent with the two-phase inside-out formation scenario.
\item The pixels outlying in the pCMDs reveal obvious features of tidal interactions in their spatial distributions. Whereas the {\bcga} appears to have experienced considerable tidal events recently, the {\bcgb} has a compact ($\sim1.4$ kpc in diameter) core with color asymmetry that possibly resulted from major dry merger at an early epoch.  
\item In the pixel luminosity functions, the {\bcgb} shows a very large faint-to-bright pixel number ratio, whereas the ratio for the {\bcga} is not so distinct from typical early-type non-BCGs. This implies that the outer body of the {\bcgb} is better evolved, probably through more minor mergers with satellite galaxies.
\end{enumerate}

These results are consistent with a picture that the BCG in the dynamically young cluster A1139 is dynamically younger and less evolved than the BCG in the relaxed cluster A2589. 
The {\bcgb} seems to have formed a larger central body from the beginning and to evolve into finally a more massive BCG through subsequent and numerous minor mergers.
Therefore, we conclude that the BCGs in A1139 and A2589 provide hints of the coevolution between BCGs and host clusters, in the context that the early and final masses as well as the dynamical state of a BCG may be closely related with the dynamical state of its host cluster.
However, this conclusion is based on the results from a case study comparing only two BCGs. To check whether our conclusion is generally established for most clusters and to understand which properties of host clusters drive the BCG evolution in detail, further investigations using a larger sample of BCGs hosted by clusters in various evolutionary stages are required.

\acknowledgments

We appreciate the anonymous referee, who motivated us to find and correct a mistake in our uncertainty measurement.
This work has used the data obtained under the K-GMT Science Program funded through Korea GMT Project operated by Korea Astronomy and Space Science Institute (KASI).
H.J.~acknowledges support from the Basic Science Research Program through the National Research Foundation of Korea (NRF), funded by the Ministry of Education (NRF-2013R1A6A3A04064993).
S.K.Y.~acknowledges support from the Korean National Research Foundation (NRF-2014R1A2A1A01003730).

\end{document}